\documentclass[a4paper,11pt]{article}

\usepackage{amsmath, amssymb, bm}
\usepackage[]{graphicx}
\usepackage[font=scriptsize,labelfont=bf]{caption}
\usepackage[caption=false]{subfig}
\usepackage{slashed}
\usepackage[top=1in, bottom=1.2in, left=0.7in, right=0.7in]{geometry}
\usepackage{authblk}
\usepackage{hyperref}
\newcommand{\be}{\begin{eqnarray}}
\newcommand{\ee}{\end{eqnarray}}

\newcommand{\bs}{\boldsymbol}

\begin{document}

\title{\bf{\Large{Milli-charged fermion vacuum polarization in cosmic magnetic fields and generation of CMB elliptic polarization}}}
\author{Damian Ejlli}

\affil{\emph{\normalsize{Department of Physics, Novosibirsk State University, Novosibirsk 630090, Russia}}}

\date{}

\maketitle

\begin{abstract}
The contribution of one loop milli-charged fermion vacuum polarization in cosmic magnetic field to the cosmic microwave background (CMB) polarization is considered. Exact and perturbative solutions of the density matrix equations of motion in terms of the Stokes parameters are presented. For linearly polarized CMB at decoupling time, it is shown that propagation of CMB photons in cosmic magnetic field(s) would generate elliptic polarization (circular and linear) of the CMB due to milli-charged fermion vacuum polarization. Analytic  expressions for the degree of circular polarization and the rotation angle of polarization plane of the CMB are presented. Depending on the ratio of the milli-charged fermion relative charge to mass, $\epsilon/m_\epsilon$, magnetic field amplitude and CMB observation frequency, it is shown that the acquired CMB degree of circular polarization could be of the order of magnitude $P_C(T_0)\sim 10^{-10}- 10^{-6}$ in the best scenario for a canonical value of the magnetic field amplitude of the order $\sim$ nG and $\epsilon/m_\epsilon\sim 10^{-4}-\text{few}\times 10^{-3}$. The effect studied also generates CMB polarization even in the case when the CMB is initially in thermal equilibrium. Limits on the magnetic field amplitude due to prior-decoupling CMB polarization are presented.

\end{abstract}


\vspace{1cm}

\section{Introduction}
 \label{sec:1}

In the standard model of particle physics is quite intriguing the fact that all known particles seem to have an electric charge that is a multiple integer of the electric charge of the $d$ quark, namely $e/3$ where $e$ is the electron charge. Even though the charge quantization apparently seems to be a fundamental principle, it is theoretically possible to have particles with electric charge $\epsilon e$ where $\epsilon$ is any real number. This possibility on the other hand is enforced by the fact that standard model of particle physics does not necessarily impose charge quantization \cite{Foot:1990mn} and charge quantization is an ad hoc assumption. In principle $\epsilon$ could be any real number but in this work I will consider only the case when $\epsilon<1$ and the particles satisfying this condition are usually called milli-charged particles.

There are essentially three ways to introduce milli-charged particles which either requires to go beyond the standard model of particle physics or stay within the standard model. Within the standard model, milli-charged particles appear by allowing the neutrino to have an electric charge which is achieved with the introduction of a right handed neutrino and by redefining the hypercharge operator. Indeed, as shown in Ref. \cite{Foot:1992ui} one can redefine the hypercharge operator as $Y\rightarrow Y=Y_\text{SM}+\epsilon(B-L)$, where $B$ and $L$ are the baryon and lepton numbers of the $U(1)$ global symmetry, and preserve the anomaly cancellation of the standard model. Another possibility to introduce milli-charged particles, is to make use of mirror symmetry, where due to mixing of photons with mirror or dark photons, charged particles under the mirror gauge group $U^\prime(1)$ couple to photons with small electric charge \cite{Holdom:1985ag}. A third possibility of manifestation of milli-charged particles appears in non abelian gauge theories with massive photons and electric charge non quantization \cite{Ignatiev:1978xj}.

So far there have been several attempts to look for milli-charged particles either directly or indirectly. From the experimental side, laser experiments are one of the main ways to look for milli-charged particles and also very weakly interacting particles such as scalar bosons, pseudoscalar particles, particles from the dark sector etc. In these experiments light is sent through an external magnetic field (usually a transverse field) and after one looks for changes in the polarization state of the incident light.
Due to the fact that there has not been any detection of milli-charged particles so far, experiments such as PVLAS  \cite{DellaValle:2015xxa} and BRFT \cite{Gies:2006ca} only put limits on the relative charge $\epsilon$ and on the milli-charged particle mass $m_\epsilon$. On the other hand, experiments which do not make use of interaction of light with an external magnetic field are those which look for invisible decay of Othopositronium \cite{Mitsui:1993ha}, Lamb shift of hydrogen atom \cite{Lundeen:1981zz}, beam dump experiment conducted at SLAC see Ref. \cite{Davidson:1991si} for more details and most recent proposed experiment to be conducted at LHC \cite{Haas:2014dda}. From the invisible decay of Orthopositronium one gets a limit on $\epsilon$ of the order $\epsilon<10^{-4}$ for $m_\epsilon<m_e$, while from beam dumb experiment such limits are by a factor two weaker. In the case of experiment to be performed at LHC \cite{Haas:2014dda}, one expects to probe directly and model independent the parameter space $10^{-3}\leq\epsilon\leq 10^{-1}$ for $0.1\,\text{GeV}\leq m_\epsilon\leq 100$ GeV \cite{Haas:2014dda}. From indirect observations, usually one gets stronger limits on $\epsilon$ from astrophysical and cosmological considerations. Typically, the tightest constraints come from big bang nucleosynthesis (BBN) with $\epsilon\lesssim \text{few}\times 10^{-9}$ for $m_\epsilon<m_e$ \cite{Mohapatra:1990vq} and from stellar evolution with $\epsilon<2\times 10^{-4}$, see Ref. \cite{Davidson:2000hf} for a review on bounds of milli-charged particles and Ref. \cite{Dolgov:2013una} for CMB bounds on abundance of milli-charged particles.

The polarization effects which laser experiments mentioned above aims to look for, would arise as consequence of interaction of the incident electromagnetic wave with the external magnetic field, where the vacuum is expected to acquire polarization due to appearance of milli-charged particles. Consequently, two polarization effects would manifest, birefringence and dichroism of the incident light. Birefringence effect is responsible for generating phase shift between the two polarization states of light while dichroism effect is responsible for changing the intensity of the incident light.

The vacuum polarization due to milli-charged particles can have wider applications especially in cosmology in the context of the CMB physics. In fact, it is well known that cosmic magnetic field(s) might have been formed in the early universe due to several mechanisms, see Ref. \cite{Grasso:2000wj} for details, and interactions of the CMB photons with such external field would represent an ideal condition for vacuum polarization to occur. In a previous work \cite{Ejlli:2016avx}, I studied the most important magneto-optic effects, including standard vacuum polarization due to electron/positron pair formation, and their impact on generation of CMB polarization. However, there are several essential differences between standard vacuum polarization and vacuum polarization due to milli-charged particles which are intrinsically related with incident photon energy, magnetic field strength and milli-charged particle mass. 

First difference is that the magnitude of birefringence and dichroism effects due to milli-charged particle vacuum polarization can be several orders of magnitude bigger than standard one. In fact, as I will show in this work, quantities of interest such as the degree of circular polarization and/or rotation angle of polarization plane of CMB, turn out to be proportional to some power of the ratio $\epsilon/m_\epsilon$ which can have large value depending on $m_\epsilon$ and $\epsilon$. Second difference is related with CMB photon energies at post decoupling epoch (that is the cosmological period which I mostly focus on) and mass of milli-charged particles. Indeed, for standard vacuum polarization, only birefringence effect would  manifest at post decoupling epoch with no dichroism effect since observed CMB photon energies are much smaller than electron mass and consequently pair production of electron/positron does not occur. But, in the case of milli-charged particles, their allowed mass range can be much smaller than CMB photon energies and consequently pair production of milli-charged particles can occur.

 In this work I study the effect of milli-charged particle vacuum polarization in cosmic magnetic field(s) on generation of mainly CMB post decoupling polarization, with emphasis on the circular polarization and on the rotation angle of the CMB polarization plane. The generation and evolution of CMB linear polarization (E-modes and B-modes) due to milli-charged vacuum polarization is not studied. In this work I assume that cosmic magnetic field has a primordial origin and consider \emph{only} fermion milli-charged particles. The cosmic magnetic field amplitude is assumed to be non stationary in time and its variation in space to be on much larger scales than CMB photon and milli-charged fermion Compton wavelengths. So, the magnetic field amplitude must be slowly varying function in space \emph{with respect} to CMB photon and milli-charged fermion Compton wavelengths. The appropriate validity of our formalism and nature of cosmic magnetic field will be defined in the subsequent sections. This paper is organized as follows: In Sec. \ref{sec:2}, I start with the photon wave equation which describes  propagation of CMB photons in non relativistic magnetized plasma in an expanding universe and derive the equations of motion for the system density matrix in terms of the Stokes parameters. Here I also outline the perturbative procedure which is used in subsequent sections. In Sec. \ref{sec:3}, I present all relevant quantities related to usual vacuum polarization and their transformation into quantities related to vacuum polarization due to milli-charged fermions. In Sec. \ref{sec:4}, I find exact solutions of equations of motion of the Stokes parameters in the case for photon propagation perpendicular with respect to external magnetic field. In this section I calculate the expected CMB degree of circular polarization at present. In Sec. \ref{sec:5}, I study the case when CMB photons propagate non orthogonal to external magnetic field and find perturbative solutions for the Stokes parameters. In Sec. \ref{sec:6}, I conclude. In this work I use the metric with signature $\eta_{\mu\nu}=\text{diag}(1, -1, -1, -1)$ and work with the natural (rationalized) Lorentz-Heaviside units ($c=k_B=\hbar=\varepsilon_0=\mu_0=1$) with $e^2=4\pi\alpha$.



\section{Equations of motion of the Stokes parameters}
\label{sec:2}
 
 In this section we focus on the equations of motion of the photon field in background magnetic field and derive the equations of motion for the Stokes parameters $I, Q, U$ and $V$ in expanding universe. Consider the case when photons propagate in a magnetized medium along the observer's $\hat{\bs n}=\hat{\bs z}$ axis and let be $\bs k=(0, 0, k)$ the photon wave vector and $\bs B_e=\left[B_e\sin(\Phi), 0, B_e\cos(\Phi)\right]$ the external magnetic field vector, where here $\Phi$ is the angle between photon direction of propagation and external magnetic field direction. The wave equation for the photon field in the FRW metric and in the WKB approximation is given by \cite{Ejlli:2016avx}
 \begin{equation}\label{eq-motion}
i\partial_t\,\Psi(k, t)=\left(M(k, B_e, \Phi)-\frac{3}{2}\,i H\,\bs I\right) \Psi(k, t),
\end{equation}
where $\Psi(k, t)=(A_+, A_\times)^\text{T}$ is a two component field, $A_+$ and $A_\times$ are respectively  the photon states perpendicular and parallel to the transverse part of the external magnetic field $\bs B_e$, $\bs I$ is the identity matrix, $H=H(t)$ is the Hubble parameter and $M$ is the mixing matrix of photon fields which does not explicitly depend on time. The mixing matrix $M$ is given by 
\begin{equation}\nonumber
M=\left(\begin{matrix}
   k-M_{+} & -iM_F \\
  iM_F  &   k-M_{\times} \\
   \end{matrix}\right), 
\end{equation} 
where $M_+=-\Pi^{22}/(2\omega)$, $M_\times=-\Pi^{11}/(2\omega)$, $iM_F=-\Pi^{12}/(2\omega)$, $\omega$ is the photon energy and $\Pi^{11}, \Pi^{22}, \Pi^{12}$ are the elements of the photon polarization tensor in magnetized medium. Here the term $iM_F$ corresponds to the Faraday effect in medium which is responsible for mixing of the photon states and the diagonal elements of the photon polarization tensor take into account birefringence and dichroism effects of photons in magnetized medium.
 
One important property of the mixing matrix $M$ is that in case when $M=M^\dagger$ the number of photons is conserved, while for $M\neq M^\dagger$ the number of photons is not conserved. In this work we consider the second possibility, namely that particle number is not conserved as it will be more clear in what follows. Since our goal is to study the effect of the cosmic magnetic field on the CMB polarization, it is more convenient to work with the Stokes parameters rather than wave equation \eqref{eq-motion}. Moreover, since the CMB is almost unpolarized, the mixing and damping of photons states during the universe evolution is better described in terms of the density matrix which satisfies the von-Neumann equation i$\partial_t\rho=[M, \rho]$. As discussed in details in Ref. \cite{Ejlli:2016avx}, the density matrix satisfies the following differential equation 
 \begin{equation}\label{dens-eq}
\frac{\partial\rho}{\partial t}=-i [\tilde M, \rho]- \{D, \rho\},
\end{equation}
 where $\rho$ is the photon polarization density matrix, $\tilde M$ is a matrix which has for diagonal elements the real part of diagonal elements of $M$ and same off diagonal elements as $M$. Here $D$ is a matrix which takes into account damping of photon field in medium and external fields. In case of photons propagating in an expanding universe, the damping matrix is composed of two terms where one term corresponds to damping of photon field in an expanding universe due to the Hubble friction and the other term takes into account decay of photons into other particles. 

 In order to make things more clear we write the diagonal elements of the matrix $M$ as, $M_+=\tilde M_{0}^{(+)}+iM_{1}^{(+)}$ and $M_\times=\tilde M_{0}^{(\times)}+iM_{1}^{(\times)}$, where the real parts of $M_+, M_\times$ in our case take into account forward scattering of photons in medium or birefringence effect while the imaginary parts take into account dichroism effect. With this splitting, the damping matrix has the form $ D=(3/2) H\bs I_{2\times 2}+\text{diag}[M_1^{(+)}, M_1^{(\times)}].$ At this point, it is convenient to express the photon polarization density matrix in terms of the Stokes parameters as shown in Ref. \cite{Fano:1954zza}, see also Ref. \cite{Ejlli:2016avx}. In this case the equations of motion for the density matrix, Eq. \eqref{dens-eq}, in terms of the Stokes parameters become
\begin{align}\label{Stokes-eq}
\dot I & =-\Delta M_1 I-\Delta M_0 Q-3H I,\nonumber\\
\dot Q & =- \Delta M_0 I-\Delta M_1 Q-2M_F U-3HQ,\\
\dot U & =2M_F Q-\Delta M_1 U+\Delta \tilde M V-3HU,\nonumber\\
\dot V &=-\Delta \tilde M U-\Delta M_1 V-3H V,\nonumber
\end{align}
where we have defined $\Delta M_0\equiv M_1^{(+)}-M_1^{(\times)}$, $\Delta M_1 \equiv M_1^{(+)}+M_1^{(\times)}$ and $\Delta\tilde M \equiv \tilde M_0^{(+)}-\tilde M_0^{(\times)}$ with the dot sign above Stokes parameters indicating the time derivative with respect to the cosmological time $t$.
 
 The linear system of Eqs. \eqref{Stokes-eq} is very general since it includes both forward scattering and decay/absorption of photons in magnetized media. It can be written in more compact form as $\dot S(t)=A(t)S(t)$ where $S=(I, Q, U, V)^\text{T}$ is the Stokes vector and $A(t)$ is the time dependent coefficient matrix which is given by
\begin{equation}\nonumber
A(t)=\left(\begin{matrix}
   -\Delta M_{1}-3H & -\Delta M_0 & 0 & 0 \\
  -\Delta M_0  &   -\Delta M_1-3H & -2 M_F & 0 \\
  0 & 2M_F & -\Delta M_1-3H & \Delta\tilde M\\
  0 & 0 & -\Delta \tilde M & -\Delta M_1-3H\\
   \end{matrix}\right).
\end{equation} 
In an expanding universe, most of quantities that enter Eqs. \eqref{Stokes-eq} have simpler form if expressed in terms of the temperature $T$ rather than $t$. Therefore, in this work we adopt this form and write the time derivative in Eqs. \eqref{Stokes-eq} as $\partial_t=-HT\partial_T$. Then we split the coefficient matrix, now as a function of the temperature, as $A(T)=B(T)+C(T)$ and write the equation for the Stokes vector as\footnote{Each Stokes parameter, in addition to the temperature $T$, depends also on the direction of propagation of CMB photons $\hat{\bs n}$ and on the angle $\Phi$. However, in order to simplify our notations, the dependence on $\hat{\bs n}$ and $\Phi$ will be omitted.}
\begin{equation}\label{eq-stokes-vec}
S^\prime(T)=[B(T)+C(T)]S(T),
\end{equation}
where $C(T)=\left[\Delta M_1/(HT)+3/T\right]\bs I_{4\times 4}$, the sign $(^\prime)$ denotes derivative with respect to $T$ and $B(T)$ is given by
\begin{equation}\nonumber
B(T)=\frac{1}{HT}\left(\begin{matrix}
   0 & \Delta M_0 & 0 & 0 \\
  \Delta M_0  &   0 & 2 M_F & 0 \\
  0 & -2M_F & 0 & -\Delta\tilde M\\
  0 & 0 & \Delta \tilde M & 0\\
   \end{matrix}\right).
\end{equation}

 In general there are not closed solutions for Eqs. \eqref{eq-stokes-vec}. This is quite common since we are dealing with first order system of differential equations with variable coefficients. In this work we look for perturbative solutions of Eqs. \eqref{eq-stokes-vec} by using regular perturbation theory. The goal is to find a reasonable splitting of the non diagonal matrix $B(T)$ in such a way that $B(T)=B_0(T)+\lambda B_1(T)$ where the parameter\footnote{Here the role of the parameter $\lambda$ is quite formal since it is not necessary to know its expression and its only purpose is to tell which matrix is considered as perturbation matrix. What we are actually requiring is to find a splitting in such a way that magnitude of elements of the perturbation matrix $\lambda B_1(T)$ are much smaller than elements of matrix $B_0(T)$. } $\lambda$ is positive and small in some sense. Therefore, we look for solution of the Stokes vector in the following form
 \begin{equation}\label{stokes-expa}
S(T)=S_0(T)+\lambda S_1(T)+\lambda^2 S_2(T)+...,
\end{equation}
where we consider the expansion up to the second order in $\lambda$. Now using expansion \eqref{stokes-expa} together with $B(T)=B_0(T)+\lambda B_1(T)$ in Eq. \eqref{eq-stokes-vec} and collecting terms with appropriate power in $\lambda$, we get the following matrix equations 
 \begin{align}\label{perturbative-sol}
S_0^\prime(T) &= \left[B_0(T)+C(T)\right]S_0(T)\nonumber,\\
\lambda S_1^\prime(T) &=  \left[B_0(T)+C(T)\right]\lambda S_1(T)+\lambda B_1(T)S_0(T),\\
\lambda^2 S_2^\prime(T) &=  \left[B_0(T)+C(T)\right]\lambda^2\,S_2(T)+\lambda B_1(T)\lambda S_1(T).\nonumber
\end{align}
Solutions of matrix equations \eqref{perturbative-sol} might be quite involved if one chooses the wrong way to split the matrix $B(T)$. The key point in order to solve Eqs. \eqref{perturbative-sol}, is to find a splitting for $B(T)$ in such a way that solutions of homogeneous equations in \eqref{perturbative-sol} are given by matrix exponential. In the next sections we deal with this problem and under what circumstances one can perform an educated splitting of $B(T)$.

 \section{One loop milli-charged fermion vacuum polarization}
 \label{sec:3}
 
As mentioned in Sec. \ref{sec:1}, in this work we study the consequences of one loop milli-charged fermion vacuum polarization to the CMB polarization.  When electromagnetic radiation interacts with an external magnetic field, is generally expected that the vacuum gets polarized as consequence of this interaction. Usually this interaction is described by the Euler-Heisenberg Lagrangian density \cite{Heisenberg:1935qt}, which takes into account non linear effects of quantum electrodynamics (QED) and consequently classical Maxwell equations are modified in order to include these new effects. For a review on Euler-Heisenberg Lagrangians see Ref. \cite{Dunne:2004nc}. At one loop level one can have either spinor or scalar contribution to vacuum polarization where their respective actions are given by $S_\text{spinor}^{(1)}=-i \ln \text{det}(i\slashed D -m_e)$ where $\slashed D$ is the Dirac operator for a given classical background photon field with $m_e$ being the electron mass and $S_\text{scalar}^{(1)}=(i/2) \ln \text{det}(D_\mu^2 +m_e^2)$.
 
 In the case of spinor QED, at one loop, it has been studied the contribution to vacuum polarization mostly due to electron/positron pair in a constant electromagnetic field\footnote{More generally the case of constant electromagnetic field can be extended to those fields whose variations over the Compton wavelength of the electron, $\lambda_C=2\pi/m_e$, and over the corresponding time interval $\tau=\lambda_C$ are much smaller than the field itself \cite{Dunne:2004nc}-\cite{BialynickaBirula:1970vy}, namely $|\partial_\mu F_{\sigma\rho}| \ll m_e\,|F_{\sigma\rho}|$ where $F_{\sigma\rho}$ is the external electromagnetic field tensor, see Sec. 2.3 of Ref. \cite{Dunne:2004nc} for details. In the case of milli-charged fermions which we study in this work, one must replace $m_e$ with $m_\epsilon$ in $|\partial_\mu F_{\sigma\rho}| \ll m_e\,|F_{\sigma\rho}|$.}. If one assumes the external field to be purely magnetic one derives the low energy limit Euler-Heisenberg Lagrangian\footnote{This Lagrangian density is obtained by expanding the full Euler-Lagrangian density for low photon energies, $\omega\ll m_e$ and magnetic field amplitude $B\ll B_c$ where $B_c=m_e^2/e$ is the critical magnetic field. } which describes light-light scattering at one loop. This scattering occurs among free photons or photons and external electromagnetic field. Although the Euler-Heinsenberg Lagrangian correctly predicts quantities related to vacuum polarization such as for example the index of refraction of light in external magnetic field, in general the process of vacuum polarization is best descibed by Schwinger proper time method \cite{Schwinger1951}.

 \begin{figure}[htbp]
\begin{center}
\includegraphics[scale=0.7]{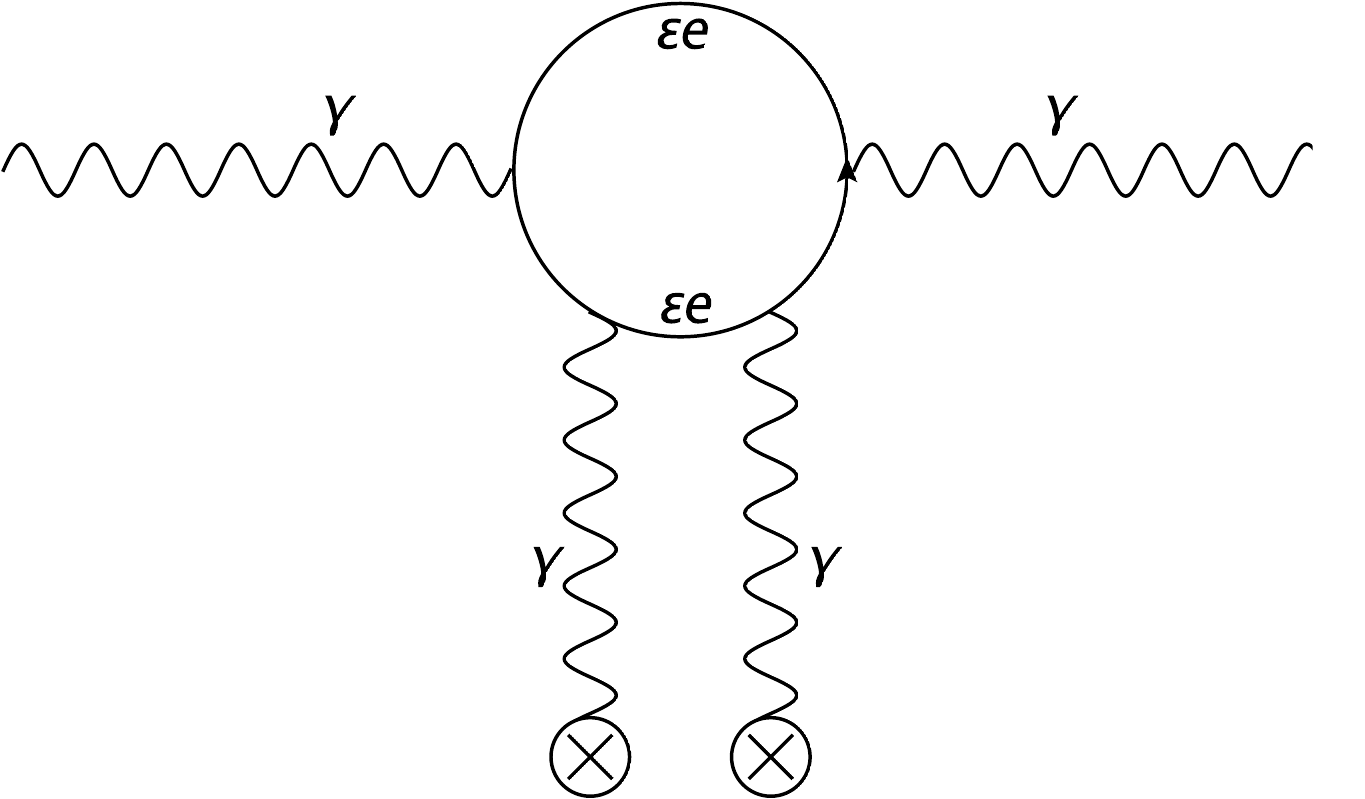}
\caption{Milli-charged fermion vacuum polarization in external electromagnetic field. The external electromagnetic field in our case is purely magnetic and is denoted with cross symbols. The fermion loop is made of particle/antiparticle milli-charged fermions with charge $Q_\epsilon=\epsilon e$ with $e$ being the electron charge and $\epsilon$ is the relative charge.}
\label{fig:milli-loop}
\end{center}
\end{figure}

One can include the effects of vacuum polarization by adding to the standard free Maxwell Lagrangian a term $-(1/2)\int d^4 x^\prime A_\mu(x)\Pi^{\mu\nu}(x, x^\prime)A_\nu(x^\prime)$ where $\Pi^{\mu\nu}$ is the photon polarization tensor in medium. Its calculation in an external electromagnetic field for spinor QED and related quantities such as the indexes of refraction and absorption have been calculated by several authors, see for example Ref. \cite{Tsai:1974fa} and references therein. In principle, also milli-charged fermion pair can contribute to $\Pi^{\mu\nu}$ in analogous way as electron/positron pair, see Fig. \ref{fig:milli-loop}. So, one would encounter the same processes that manifest in the case of vacuum polarization due to electron/positron pair but with different magnitude depending on milli-chrged fermion mass $m_\epsilon$ and $\epsilon$.

Depending on the incident photon energy $\omega$ and strength of external magnetic $B_e$, one can have birefringence and/or dichroism effects due to vacuum polarization.  As already mentioned, in this work we are interested to investigate these effects in the case of vacuum polarization due to milli-charged fermions. This can be done by using the results obtained in the case of vacuum polarization due to electron/positron pair and adopt them to our case, by simply doing the following identifications, $e\rightarrow \epsilon e$ and $m_e\rightarrow m_\epsilon$. We can use the results found in Ref. \cite{Tsai:1974fa} in the case vacuum polarization due to  electron/positron pair for $B_e\ll B_c$ and adopt them to the case of milli-charged fermion vacuum polarization for\footnote{The case when $\epsilon(m_e/m_\epsilon)^2B_e\sin(\Phi)\gg B_c$ is not studied in this work.}, $\epsilon(m_e/m_\epsilon)^2B_e\ll B_c$.   Therefore we get
\begin{equation}\label{refrec-index}
\Delta\tilde M=-\left(\frac{\epsilon\,m_e}{m_\epsilon}\right)^4\omega\left(\frac{\alpha}{4\pi}\right)\left(\frac{B_e}{B_c}\right)^2\sin^2(\Phi)\Delta\mathcal I(\chi),
\end{equation}
where $\Delta\tilde M=\left(\text{Re}\{\Pi^{11}\}-\text{Re}\{\Pi^{22}\}\right)/(2\omega)=\omega(n_+-n_\times)$ with $n_+, n_\times$ being respectively photon indexes of refraction for the states $A_+, A_\times$. Here $\Delta\mathcal I(\chi)$ is given by
\begin{equation}\nonumber
\Delta\mathcal I(\chi)=-2^{-2/3}(3/\chi)^{4/3}\int_0^1 dv\, (1-v^2)^{2/3}\,\tilde e_0^\prime\left[-\left(\frac{6}{\chi(1-v^2)}\right)^{2/3}\right]
\end{equation}
where $\tilde e_0(y)\equiv \int_0^\infty dx \sin\left(xy-x^3/3\right)$ is a kind of generalized Airy function with $\tilde e_0^\prime=d\tilde e_0/dy$ and $\chi$ is a parameter which is defined as 
\begin{equation}\nonumber
\chi\equiv\frac{3}{2}\epsilon\left(\frac{\omega}{m_e}\right)\left(\frac{m_e}{m_\epsilon}\right)^3\left(\frac{B_e}{B_c}\right)\sin(\Phi).
\end{equation}
The function $\Delta\mathcal I(\chi)$ has the following asymptotic expressions in the case when $\chi\ll1$ and $\chi\gg 1$
 \begin{equation}\label{asympto-I}
\Delta\mathcal I(\chi)=6/45\quad (\chi\ll 1), \qquad \Delta\mathcal I(\chi)=-\frac{9}{7}\sqrt{\pi}\,2^{1/3}\frac{\Gamma^2(2/3)}{\Gamma(1/6)}\,\chi^{-4/3}\qquad (\chi\gg 1).
\end{equation}

 Expression \eqref{refrec-index} takes into account forward scattering of photons (index of refraction) in the presence of milli-charged fermion loop for $\epsilon(m_e/m_\epsilon)^2B_e\ll B_c$. On the other hand, if photon energy is $\omega>2 m_\epsilon$, also take place pair production of milli-charged fermions in external magnetic field and consequently, for each photon state there is also an absorption/decay index. Absorption processes are encoded in the imaginary part of the photon polarization tensor, where in general the absorption coefficients are related to the polarization tensor through $\kappa_{+}=-(1/\omega)\text{Im}\{\Pi^{22}\}$ and $\kappa_{\times}=-(1/\omega)\text{Im}\{\Pi^{11}\}$ . Using expressions for absorption/decay coefficients $\kappa_+$ and $\kappa_\times$ derived in Ref.  \cite{Tsai:1974fa} for initially polarized light and for $\epsilon(m_e/m_\epsilon)^2B_e \ll B_c$, we find the following expressions for $\Delta M_0$ and $\Delta M_1$
 \begin{align}
\Delta M_0 &=\kappa_+-\kappa_\times=\frac{1}{2}\epsilon^3\left(\frac{m_e}{m_\epsilon}\right)\alpha\,\omega_c\,\sin(\Phi)\Delta\mathcal T_0(\chi),\nonumber\\
\Delta M_1 & =\kappa_++\kappa_\times=\frac{1}{2}\epsilon^3\left(\frac{m_e}{m_\epsilon}\right)\alpha\,\omega_c\,\sin(\Phi)\Delta\mathcal T_1(\chi)\nonumber,
\end{align}
 where $\omega_c=eB_e/m_e$ is the cyclotron frequency and $\Delta\mathcal T_{0, 1}(\chi)$ are respectively given by
 \begin{equation}
\Delta\mathcal T_0(\chi) =-\frac{2\sqrt{3}}{\pi\chi}\int_0^1 dv\,K_{2/3}\left[\frac{4}{\chi(1-v^2)}\right],\quad
\Delta\mathcal T_1(\chi) =\frac{2\sqrt{3}}{\pi\chi}\int_0^1 dv\, \frac{\left({3}-\frac{v^2}{3}\right)}{(1-v^2)}\,K_{2/3}\left[\frac{4}{\chi(1-v^2)}\right],
\end{equation}
where $K_{2/3}$ is the modified Bessel function of the second kind or the so called MacDonald function. In the cases when $\chi\ll 1$ and $\chi\gg 1$, the  functions $\Delta\mathcal T_0$ and $\Delta\mathcal T_1$ have the following asymptotic expressions
\begin{align}\label{DI-0}
\Delta\mathcal T_0(\chi) &=-\frac{1}{4}\sqrt{\frac{3}{2}}e^{-4/\chi} \quad (\chi\ll 1), \qquad \Delta\mathcal T_0(\chi)=-\frac{2^{1/3}\sqrt{3}\,\Gamma^2(2/3)}{7\sqrt{\pi}\,\Gamma(7/6)}\chi^{-1/3}\quad (\chi\gg 1),\nonumber\\
\Delta\mathcal T_1(\chi) &=\frac{3}{4}\sqrt{\frac{3}{2}}e^{-4/\chi} \quad (\chi\ll 1), \qquad \Delta\mathcal T_1(\chi)=\frac{2^{1/3}\,5\,\sqrt{3}\,\Gamma^2(2/3)}{7\sqrt{\pi}\,\Gamma(7/6)}\chi^{-1/3}\quad (\chi\gg 1).
\end{align}

\section{Solutions of equations of motion in case of $\Phi=\pi/2$.}
\label{sec:4}
 
 In this section we concentrate on the solution of equation of the Stokes vector in the particular case when $\Phi=\pi/2$, where in this work we consider for simplicity $\Phi$ in the interval $0\leq \Phi\leq \pi/2$. This case corresponds to light propagation perpendicular to the external magnetic field where the Faraday effect is completely absent since this effect is proportional to $\cos(\Phi)$. For this particular case it is not necessary to use perturbation theory since one can find exact solution for the Stokes vector. In absence of the Faraday effect, in matrix $B(T)$ enters $\Delta M_0$ which includes the dichroism effect caused by absorption/decay and consequently appearance of real milli-charged fermions for photon energies $\omega>2m_\epsilon$ and $\Delta\tilde M$ which includes only birefringence effect caused by virtual appearance of milli-charged fermions in external magnetic field.


 Before proceeding to the solution of the Stokes vector, it is very important to make some general discussions on values of the parameters $\epsilon$ and $m_\epsilon$ which we consider in this work. The first thing is about expressions $\Delta\mathcal T_{0, 1}$ in \eqref{DI-0}. They have been derived in the case when $\omega\geq 2m_\epsilon$ and in the case when the number of Landau levels is very large. Indeed, it is well known that in the presence of an external magnetic field, orbits of charged particles and their corresponding energies and angular momenta are quantized. Consequently, due to energy and angular momentum quantization, the incident light would manifest absorption lines for the states $A_+$ and $A_\times$ and its spectrum would acquire a sawtooth-like form. However, as shown in Ref. \cite{Daugherty:1984tr} as far as the number of Landau levels is very high, namely $N_L= (1/24)\epsilon^{-2}(\omega/m_e)^4(B_e/B_c)^{-2}\gg 1$ (where we have adopted the expression for $N_L$ to the case of milli-charged fermions), the spacing between absorption peaks is very narrow. Moreover, by averaging over small energy intervals the expressions for absorptions coefficients found in Ref. \cite{Daugherty:1984tr}, one would remove the sawtooth-like behaviour and absorption coefficients found in Ref. \cite{Daugherty:1984tr} agree with the smooth in $\omega$ asymptotic expressions found in Ref. \cite{Tsai:1974fa}. 
 
The second thing is related to the term $\Delta\tilde M$ which essentially takes into account only the birefringence effect due to appearance of virtual milli-charged fermions. We may note from expression \eqref{refrec-index} that in the case when $\epsilon\rightarrow 1$ and $m_\epsilon\rightarrow m_e$, one would get an expression which coincides exactly with that of vacuum polarization due to electron/positron pair. Since is not of particular interest the case when birefringence effect due to milli-charged fermion vacuum polarization is smaller than usual birefringence effect due to standard vacuum polarization, in this work \emph{we require} that $\epsilon m_e/m_\epsilon \geq 1$.

After these considerations, we can proceed on looking for solution of the Stokes vector in the case of absent Faraday effect. Now by setting $M_F(T)=0$ in the matrix $B(T)$ and then by noting that the commutator $[B(T_1), B(T_2)]=0$ for $T_1\neq T_2$, the solution of Eq. \eqref{eq-stokes-vec} is given by taking the exponential of $B(T)+C(T)$. Therefore we get the following exact solutions for the components of $S(T)$
\begin{align}\label{sol-no-far}
I(T) &=\frac{1}{2}\left[(\exp[-\mathcal G_0(T)]+\exp[\mathcal G_0(T)]) I_i+(\exp[-\mathcal G_0(T)]-\exp[\mathcal G_0(T)]) Q_i\right]\left(\frac{T}{T_i}\right)^2\exp[-\mathcal G_1(T)],\nonumber\\
Q(T) &=\frac{1}{2}\left[(\exp[-\mathcal G_0(T)]-\exp[\mathcal G_0(T)]) I_i+(\exp[-\mathcal G_0(T)]+\exp[\mathcal G_0(T)]) Q_i\right]\left(\frac{T}{T_i}\right)^2\exp[-\mathcal G_1(T)],\nonumber\\
U(T) &=\left(\cos[\tilde{\mathcal G}(T)] U_i+\sin[\tilde{\mathcal G}(T)] V_i\right)\left(\frac{T}{T_i}\right)^2\exp[-\mathcal G_1(T)],\nonumber\\
V(T) &=\left(-\sin[\tilde{\mathcal G}(T)] U_i+\cos[\tilde{\mathcal G}(T)] V_i\right)\left(\frac{T}{T_i}\right)^2\exp[-\mathcal G_1(T)]
\end{align}
where we have defined 
\begin{equation}
\tilde{\mathcal G}(T)\equiv \int_{T}^{T_i} \frac{\Delta\tilde M(T^\prime)}{H(T^\prime)T^\prime}\,dT^\prime, \quad {\mathcal G}_0(T)\equiv \int_{T}^{T_i} \frac{\Delta M_0(T^\prime)}{H(T^\prime)T^\prime}\,dT^\prime, \quad {\mathcal G}_1(T)\equiv \int_{T}^{T_i} \frac{\Delta M_1(T^\prime)}{H(T^\prime)T^\prime}\,dT^\prime,
\end{equation}
and $I_i=I(T_i), Q_i=Q(T_i), U_i=U(T_i), V_i=V(T_i)$. Here $T_i$ is the initial temperature of the CMB with $T\leq T_i$.

In order to apply solutions \eqref{sol-no-far} to a concrete problem is necessary to calculate the expressions for $\mathcal{\tilde G}(T)$ and $\mathcal G_0(T)$. However, is not necessary to calculate the expression for $\mathcal G_1(T)$ since it is common to all Stokes parameters and it cancels out in those expressions (which interests us) that contain their ratio. This term is important only in those situations where is required to know the magnitude of Stokes parameters as function of $T$, since it corresponds to a damping term which changes the photon number. We may also note the effective damping term $(T/T_i)^2$ due to the universe expansion, see Ref. \cite{Ejlli:2016avx} for details. Since in an expanding universe all quantities which enter in $\mathcal G_0(T)$ and $\mathcal{\tilde G}(T)$ depend on $T$, it necessary to write down their dependence. These quantities are the photon energy $\omega(T)$, external magnetic field amplitude $B_e(T)$ and the Hubble parameter $H(T)$. In an expanding universe, the photon energy scales as $\omega(T)=\omega_0(T/T_0)$, the magnetic field amplitude scales as $B_e(T)=B_{e0}(T/T_0)^2$ where for the latter expression magnetic flux  conservation in the cosmological plasma is assumed. Here $\omega_0=\omega(T_0)$ is the present value of photon energy and $B_{e0}=B_e(T_0)$ is the present value of magnetic field amplitude. On the other hand, the Hubble parameter depends on $T$ and on present values of density parameters of matter $\Omega_M$, radiation $\Omega_R$ and on that corresponding to the vacuum energy $\Omega_\Lambda$. In this work we concentrate at the post decoupling epoch where only matter and vacuum energy contribute to $H$. However, since vacuum energy contributes only for temperatures close to present value of the CMB, namely $T\sim T_0$, we neglect also its contribution to $H$. So, to good accuracy we can write the Hubble parameter with only matter contribution, namely $H(T)\simeq H_0\sqrt{\Omega_M}(T/T_0)^{3/2}$ where $H_0$  is the present value of the Hubble parameter and $h_0^2\Omega_M=0.12$ with $h_0=0.67$ according to the Planck collaboration \cite{Ade:2015xua}.

Now let us focus on first on the term $\mathcal G_0(T)$ and write it in the form 
\begin{equation}\label{G-0}
\mathcal G_0(T)=\mathcal A \int_T^{T_i} T^{\prime -1/2}\Delta \mathcal T_0[\chi(T^\prime)]\, dT^{\prime}, 
\end{equation}
where $\mathcal A$ is defined as
\begin{equation}\nonumber 
\mathcal A\equiv 2.11\times 10^{22}\,\epsilon^3\left(\frac{m_e}{m_\epsilon}\right)\left(\frac{B_{e0}}{\text{G}}\right)\sqrt{T_0}\,\sin(\Phi)\quad (\text{K}^{-1}).
\end{equation}
Moreover, since in the function $\Delta \mathcal T_0$ enters the parameter $\chi(T)$, it is convenient to write the latter as $\chi(T)=\mathcal B\,T^3$ where $\mathcal B$ is defined as
\begin{equation}\nonumber
\mathcal B\equiv 2.74\times 10^{-34}\,\epsilon\left(\frac{m_e}{m_\epsilon}\right)^3\left(\frac{B_{e0}}{\text{G}}\right)\left(\frac{\nu_0}{\text{Hz}}\right)T_0^{-3}\,\sin(\Phi),
\end{equation}
where we used $\omega_0=2\pi\nu_0$ with $\nu_0$ being the photon frequency at present. Second, the term $\tilde{\mathcal G}(T)$ can be written in the following form
\begin{equation}\label{TG}
\tilde{\mathcal G}(T)=\mathcal C\,\int_T^{T_i}\,T^{\prime 5/2}\Delta\mathcal I[\chi(T^\prime)]\,dT^\prime,
\end{equation}
where we defined $\mathcal C$ as 
\begin{equation}\nonumber
\mathcal C\equiv -6.09\times 10^{-13}\,\left(\frac{\epsilon\,m_e}{m_\epsilon}\right)^4\left(\frac{\nu_0}{\text{Hz}}\right)\left(\frac{B_{e0}}{\text{G}}\right)^2\,T_0^{-5/2}\,\sin^2(\Phi)\quad (\text{K}^{-1}).
\end{equation}

\subsection{Generation of polarization in the case of $\chi\ll 1$.}

Now let us concentrate on the generation of CMB polarization in the case when $\chi\ll 1$. By using the asymptotic expression for $\Delta\mathcal I(\chi)$ in \eqref{asympto-I}, the expression for $\tilde{\mathcal G}(T)$ in \eqref{TG} becomes
\begin{equation}\label{TG-1}
\tilde{\mathcal G}(T)=\frac{12}{315}\,\mathcal C\,(T_i^{7/2}-T^{7/2}).
\end{equation}
On the other hand for $\chi\ll1$,  by using the expression \eqref{G-0} we get
\begin{align}\label{G-0-1}
\mathcal G_0(T) &=-\frac{1}{4}\sqrt{\frac{3}{2}}\mathcal A \int_T^{T_i} T^{\prime -1/2} \exp(-4/{\mathcal B} T^{\prime 3})\, dT^{\prime}\nonumber \\
& = -\frac{1}{2}\sqrt{\frac{3}{2}}\mathcal A\left[\sqrt{T_i}\exp(-4/\mathcal BT_i^{3})- \sqrt{T}\exp(-4/\mathcal BT^{3})+2^{1/3}\mathcal B^{-1/6}\left(\Gamma\left(\frac{5}{6}, \frac{4}{\mathcal B T^3}\right)- \Gamma\left(\frac{5}{6}, \frac{4}{\mathcal B T_i^3}\right)\right)\right],
\end{align}
where we used the expression for $\Delta\mathcal I_0(\chi)$ in \eqref{DI-0} for $\chi\ll1$ and the definition of the generalized incomplete Euler gamma function, $\Gamma(s, x_1, x_2)=\Gamma(s, x_1)-\Gamma(s, x_2)$ with $\Gamma(s, x)=\int_x^\infty\, dt\,t^{s-1} e^{-t}$ being the incomplete Euler gamma function. We may note that in $\mathcal G_0$ enters the function $\Delta M_0$ which takes into account decay of photons into in milli-charged fermions. This process occurs as far as $\omega(T)\geq 2m_\epsilon$ or until when the temperature $T\geq T_\epsilon= (2m_\epsilon/\omega_0)T_0$ where $T_\epsilon$ is the minimum temperature for the decay to occur and used $\omega(T)=\omega_0(T/T_0)$. So, assuming that the decay happens for temperatures $T_\epsilon< T_i$, the actual limits of integration in \eqref{G-0} are for $T_\epsilon\leq T\leq T_i$ if $T_0< T_\epsilon$ or $T_0\leq T\leq T_i$, if $T_\epsilon< T_0$. Consequently, one must replace the lower limit of integration in \eqref{G-0-1} with $T\rightarrow T_\epsilon=(2m_\epsilon/\omega_0)T_0$ if $T_0<T_\epsilon$ or with $T\rightarrow T_0$ if $T_\epsilon<T_0$. On the other hand, having assume that the decay occurs for $T_\epsilon < T_i$, we get an additional constraint on the mass of milli-charged fermion which must be $m_\epsilon < (T_i/T_0)(\omega_0/2)$ for given $T_0, T_i$ and $\omega_0$. 

So far, there have been several constraints on $\epsilon$ and $m_\epsilon$ which we must be aware before proceeding further. The condition $\chi(T)\ll 1$ must be satisfied for all $T_0\leq T\leq T_i$ and this is fulfilled only when 
\begin{equation}\label{chi-cond}
\epsilon\left(\frac{m_e}{m_\epsilon}\right)^3\ll 3.65\times 10^{33}\left(\frac{\text{G}}{B_{e0}}\right)\left(\frac{\text{Hz}}{\nu_0}\right)\left(\frac{T_0}{T_i}\right)^3.
\end{equation}
There are also the conditions, $\epsilon m_e/m_\epsilon\geq 1$ and $m_\epsilon< (T_i/T_0)(\omega_0/2)$ which can be written respectively in the following form
\begin{equation}\label{qed-dec-cond}
\left(\frac{m_\epsilon}{m_e}\right)\leq \epsilon<1, \quad \left(\frac{m_\epsilon}{\text{eV}}\right)< 2.06\times 10^{-15}\left(\frac{T_i}{T_0}\right)\left(\frac{\nu_0}{\text{Hz}}\right),
\end{equation}
where we putted the condition $\epsilon<1$ in the first expression in \eqref{qed-dec-cond} since we are dealing with milli-charged particles. The second condition in \eqref{qed-dec-cond} has the following physical interpretation: given the CMB photon observation frequency at present $\nu_0$, the condition $m_\epsilon< 2.06\times 10^{-15}(T_i/T_0)(\nu_0/\text{Hz})$ eV represents the  values of milli-charged fermion masses that CMB photons decay into for $T_\epsilon<T_i$. If $m_\epsilon\geq 2.06\times 10^{-15}(T_i/T_0)(\nu_0/\text{Hz})$ eV, it represents the lowest milli-charged fermion mass that CMB photons do not decay into for $T_\epsilon\geq T_i$. In what follows we consider $T_i$ to be the CMB temperature at post decoupling time if not otherwise specified. So, the condition $T_\epsilon<T_i$ would mean decay into milli-charged fermions at post decoupling time while $T_\epsilon\geq T_i$ would mean non decay into milli-charged fermions at post decoupling time for given observation frequency $\nu_0$.
The condition that number of Landau levels must be $N_L\gg 1$ is satisfied for all $T_0\leq T\leq T_i$ when 
\begin{equation}\label{landau-cond}
\epsilon\ll 5.95\times 10^{-28}\left(\frac{\nu_0}{\text{Hz}}\right)^2\left(\frac{\text{G}}{B_{e0}}\right).
\end{equation}
For $\nu_0\geq 10^{10}$ Hz and $B_{e0}\leq 10^{-9}$ G, the condition $N_L\gg 1$ is always satisfied for $\epsilon<1$. The last condition to be satisfied is $\epsilon(m_e/m_\epsilon)^2 B_e\ll B_c$, which is fulfilled for all $T_0\leq T\leq T_i$ only when $T=T_i$
\begin{equation}\label{critic-cond}
\epsilon\left(\frac{m_e}{m_\epsilon}\right)^2\ll 4.44\times 10^{13}\,\left(\frac{T_0}{T_i}\right)^2\left(\frac{\text{G}}{B_{e0}}\right).
\end{equation}

Now with analytic expressions for $\tilde{\mathcal G}(T)$ in \eqref{TG-1} and $\mathcal G_0(T)$ in \eqref{G-0-1} and with conditions \eqref{chi-cond}-\eqref{critic-cond} we have all necessary quantities in order to treat generation of CMB polarization. Let us start first with generation of circular polarization and calculate its degree of polarization, $P_C(T)=|V(T)|/I(T)$, which at $T=T_0$ is given by
\begin{equation}\label{deg-circ-pol-0}
P_C(T_0)=\frac{2\left|-\sin[\tilde{\mathcal G}(T_0)] U_i+\cos[\tilde{\mathcal G}(T_0)] V_i\right|}{(\exp[-\mathcal G_0(T)]+\exp[\mathcal G_0(T)]) I_i+(\exp[-\mathcal G_0(T)]-\exp[\mathcal G_0(T)]) Q_i}.
\end{equation}
We may note from expression \eqref{deg-circ-pol-0} two important things. First the terms corresponding to dilution due to universe expansion $(T/T_i)^2$ and decay of photons into milli-charged fermions encoded in $\exp[-\mathcal G_1(T)]$ cancel out exactly. Second we may note that in case of unpolarized CMB at $T=T_i$, namely $Q_i=U_i=V_i=0$, there is not generation of circular polarization for $T\leq T_i$.

Obviously, in order to have generation of circular polarization, the CMB must be polarized at $T=T_i$. Since in this section we concentrate at the post decoupling epoch, we assume that the CMB acquires linear polarization at decoupling time or $T_i=2970$ K due Thomson scattering of CMB photons on electrons. However, as it is well known Thomson scattering does not generate circular polarization, so for the moment we consider $Q_i\neq 0, U_i\neq 0$ and $V_i=0$. Here $Q_i$ and $U_i$ are calculated in a common reference system \cite{Kosowsky:1994cy}.

We may note from expression \eqref{G-0-1} that in the first and second term appear $\chi(T)$ and $\chi(T_i)$ in the exponentials. However, since $\chi(T)\ll 1$ in the whole interval $T_0\leq T\leq T_i$, we have that $\chi(T)<\chi(T_i)\ll1$ since $\chi$ is proportional to $T^3$. Consequently, the first term in  \eqref{G-0-1} is bigger than the second and the fourth term is bigger than the third one. In this case we can approximate
\begin{equation}\label{G-0-2}
\mathcal G_0(T) \simeq -\frac{1}{2}\sqrt{\frac{3}{2}}\mathcal A\left[\sqrt{T_i}\exp(-4/\mathcal BT_i^{3})- 2^{1/3}\mathcal B^{-1/6} \Gamma\left(\frac{5}{6}, \frac{4}{\mathcal B T_i^3}\right)\right].
\end{equation}
From expression \eqref{G-0-2} we may note that $\exp(-4/\mathcal BT_i^{3})$ and $\Gamma\left(5/6, 4/\mathcal B T_i^3\right)$ approach to zero quite fast. Indeed for $\chi(T_i)=\mathcal B T_i=0.1$ their numerical values are respectively $\exp(-4/\mathcal BT_i^{3})=4.24\times 10^{-18}$ and $\Gamma\left(5/6, 4/\mathcal B T_i^3\right)=2.28\times 10^{-18}$. If for example $\chi(T_i)=0.01$ one would get exceedingly small values $\exp(-4/\mathcal BT_i^{3})=1.91\times 10^{-174}$ and $\Gamma\left(5/6, 4/\mathcal B T_i^3\right)=7.05\times 10^{-175}$. In the case when $\chi(T_i)\simeq 0.1$ there are some values of $\epsilon$ and $m_\epsilon$ within the constraints \eqref{chi-cond}-\eqref{critic-cond} which give values of $\mathcal A$ of the same order of magnitude or even bigger than $\exp(-4/\mathcal BT_i^{3})$ and $\Gamma\left(5/6, 4/\mathcal B T_i^3\right)$, so, $\mathcal G_0(T_0)$ can have values of order of unity or even bigger\footnote{If one considers other constraints in addition to \eqref{chi-cond}-\eqref{critic-cond}, the statement that $\mathcal G_0(T_0)$ might be of order of unity or higher might not be true.}. On the other hand, there are values of parameters $\epsilon$ and $m_\epsilon$ for which values of incomplete Gamma function and exponential term are exceedingly small, so we can approximate to high accuracy $|\mathcal G_0(T_0)|\ll 1$.

For values of $\epsilon$ and $m_\epsilon$ satisfying conditions \eqref{chi-cond}-\eqref{critic-cond} and when $|\mathcal G_0(T_0)|\ll 1$, the expression \eqref{deg-circ-pol-0} for the degree of circular polarization can be approximated as
\begin{equation}\label{deg-circ-pol-1}
P_C(T_0)\simeq \left|-\sin[\tilde{\mathcal G}(T_0)] (U_i/I_i)\right|.
\end{equation}
The expression for $\tilde{\mathcal G}(T)$ at present would be $\tilde{\mathcal G}(T_0)\simeq (12/315)\,\mathcal C\, T_i^{7/2}$ since $T_i\gg T_0$ and the expression for the degree of circular polarization \eqref{deg-circ-pol-1} becomes
\begin{equation}\label{deg-circ-pol-2}
P_C(T_0)\simeq \left| \sin\left[2.7\times 10^{-3}\,\left(\frac{\epsilon\,m_e}{m_\epsilon}\right)^4\,\left(\frac{\nu_0}{\text{Hz}}\right)\,\left(\frac{B_{e0}}{\text{G}}\right)^2\right]\frac{U_i}{I_i}\right|,
\end{equation}
where in $\mathcal C$ we took $T_i/T_0=1+z=1090$ which corresponds to the redshift of the decoupling time. In the case when the sine function in \eqref{deg-circ-pol-2} is equally to one or equivalently when 
\begin{equation}\label{unity-cond}
\epsilon\left(\frac{m_e}{m_\epsilon}\right)=4.38\,\left[2\pi n+\frac{1}{2}\right]^{1/4} \left(\frac{\text{Hz}}{\nu_0}\right)^{1/4}\,\left(\frac{\text{G}}{B_{e0}}\right)^{1/2},\quad n\geq 0 \quad (n\in \bf Z),
\end{equation}
we get the interesting situation when $P_C(T_0)\simeq |U_i/I_i|$. On the other hand, in the case when the argument of sine function in \eqref{deg-circ-pol-2} is smaller than one, we get 
\begin{equation}\label{deg-circ-pol-3}
P_C(T_0)\simeq 2.7\times 10^{-3}\,\left(\frac{\epsilon\,m_e}{m_\epsilon}\right)^4\,\left(\frac{\nu_0}{\text{Hz}}\right)\,\left(\frac{B_{e0}}{\text{G}}\right)^2\left|\frac{U_i}{I_i}\right|,
\end{equation}
where must be satisfied
\begin{equation}\label{weak-cond}
\epsilon\left(\frac{m_e}{m_\epsilon}\right)\ll 4.38 \left(\frac{\text{Hz}}{\nu_0}\right)^{1/4}\,\left(\frac{\text{G}}{B_{e0}}\right)^{1/2}.
\end{equation}

Let us take for example $B_{e0}=1$ nG and $\nu_0=50$ GHz where the condition \eqref{weak-cond} becomes $\epsilon(m_e/m_\epsilon)\ll 293$. If we take for example $\epsilon(m_e/m_\epsilon)=100$, we would get from expression \eqref{deg-circ-pol-3}, $P_C(T_0)=1.35\times 10^{-2}|U_i/I_i|$. If we take $B_{e0}=10$ nG and $\nu_0=100$ GHz, the condition \eqref{weak-cond} becomes $\epsilon(m_e/m_\epsilon)\ll 77.8$. By choosing say $\epsilon(m_e/m_\epsilon)=50$, we would get $P_C(T_0)=0.16|U_i/I_i|$.
If we take $\nu_0=10^{8}$ Hz and $B_{e0}=1$ nG, we get $\epsilon(m_e/m_\epsilon)\ll 1385$ and by taking $\epsilon (m_e/m_\epsilon)=1000$ we would get $P_C(T_0)=0.27|U_i/I_i|$ while for $\epsilon (m_e/m_\epsilon)=1200$ we would get $P_C(T_0)=0.56|U_i/I_i|$. Consequently, as far as the condition \eqref{weak-cond} is satisfied, the degree of circular polarization today is bounded between
\begin{equation}\nonumber
2.7\times 10^{-9}\,\left(\frac{\nu_0}{\text{Hz}}\right)\,\left(\frac{B_{e0}}{\text{G}}\right)^2\leq P_C(T_0)\lesssim |U_i|/I_i.
\end{equation}
It is worth to note that in the case when $m_\epsilon\geq (T_i/T_0)(\omega_0/2)$, there is not decay of CMB photons into milli-charged fermions, for given energy $\omega_0$, at the post decoupling epoch since this occurs before. In this case the term $\mathcal G_0(T)=0$ is automatically satisfied and only birefringence effect manifest at the post decoupling epoch. All previous conclusions for the value of $P_C(T_0)$ in the case $|\mathcal G_0(T_0)|\ll 1$ apply also for $\mathcal G_0(T_0)=0$. 

When $\mathcal G_0(T_0)=0$, the mass of the milli-charged fermion is bounded from below and satisfies $m_\epsilon\geq 2.06\times 10^{-15}(T_i/T_0)(\nu_0/\text{Hz})$ eV=$2.24\times 10^{-12}(\nu_0/\text{Hz})$ eV, where we used the fact that $T_i/T_0=1090$, namely the redshift of decoupling epoch. Suppose we observe the CMB at the frequencies $\nu_0=10^8$ Hz and $\nu_0=33$ GHz. For these values of the frequencies, we have respectively that $m_\epsilon\geq 2.24\times 10^{-4}$ eV and $m_\epsilon\geq 7.4\times 10^{-2}$ eV. Now if we consider the results found above such for example $\epsilon(m_e/m_\epsilon)\ll 1385$ at $\nu_0=10^8$ Hz and $B_{e0}=1$ nG and by taking $\epsilon (m_e/m_\epsilon)=1000$, we get for $m_\epsilon\geq 2.24\times 10^{-4}$ eV, $\epsilon\geq 4.4\times 10^{-7}$. If $m_\epsilon\geq 7.4\times 10^{-2}$ eV, we have for $\nu_0=33$ GHz and $B_{e0}=1$ nG, $\epsilon(m_e/m_\epsilon)\ll 325$. By taking for example $\epsilon (m_e/m_\epsilon)=250$, we get for $m_\epsilon\geq 7.4\times 10^{-2}$ eV, $\epsilon\geq 3.62\times 10^{-5}$ etc. Obviously for these values of $m_\epsilon$ and $\epsilon$, all conditions \eqref{chi-cond}, \eqref{landau-cond} and \eqref{critic-cond} are satisfied. In the case when the condition \eqref{unity-cond} is satisfied, the degree of circular polarization is equal to $|U_i/I_i|$. Now, if we consider that $m_\epsilon\geq 2.24\times 10^{-4}$ at $\nu_0=10^8$ Hz, we get for $B_{e0}=1$ nG and for $n=0$, the following lower limit $\epsilon\geq 6.83\times 10^{-7}$. Similar conclusions can be done for other values of the parameters. 

\begin{figure*}[h!]
\centering
\mbox{
\subfloat[\label{fig:Fig1}]{\includegraphics[scale=0.65]{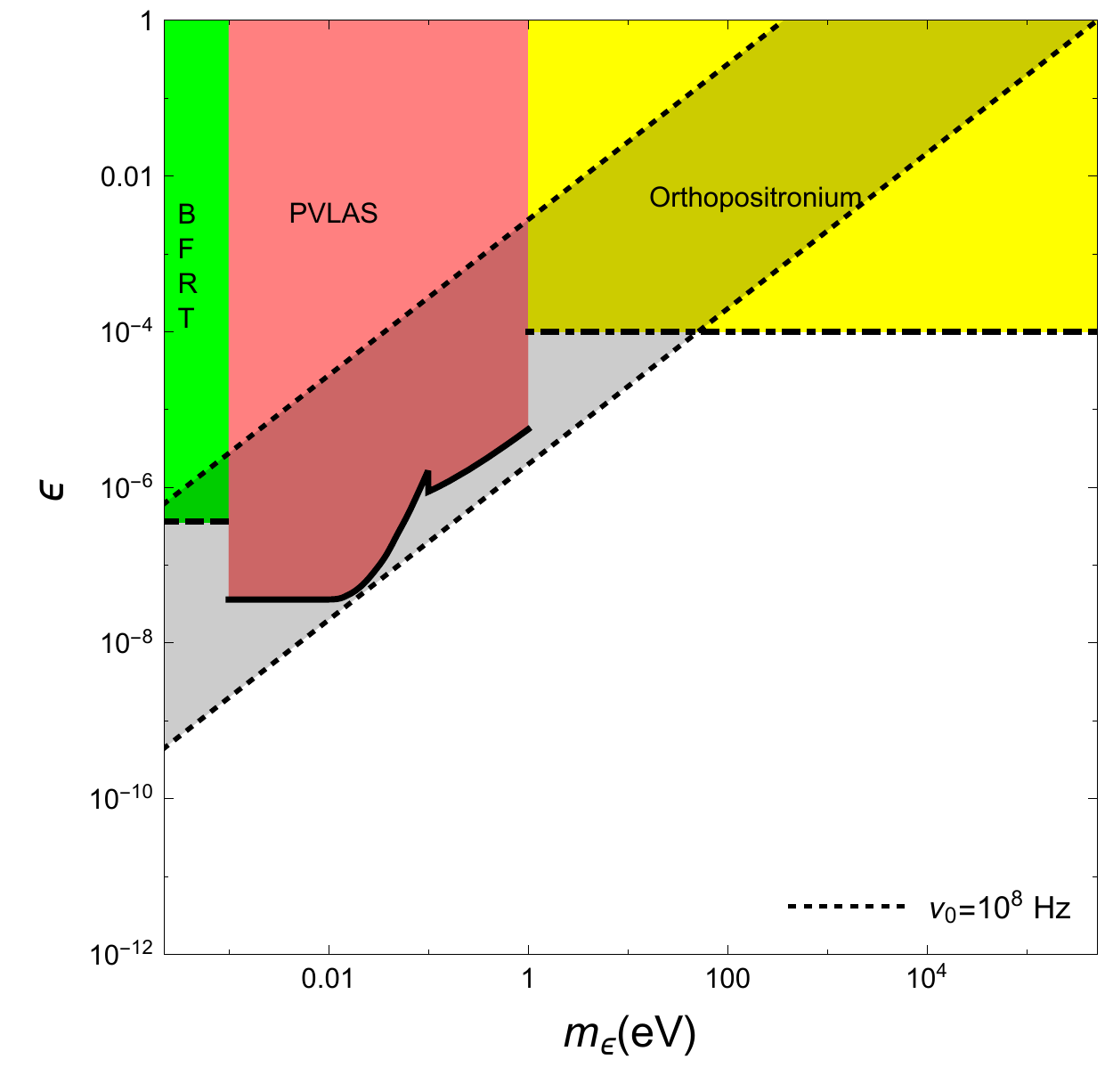}}\qquad
\subfloat[\label{fig:Fig2}]{\includegraphics[scale=0.65]{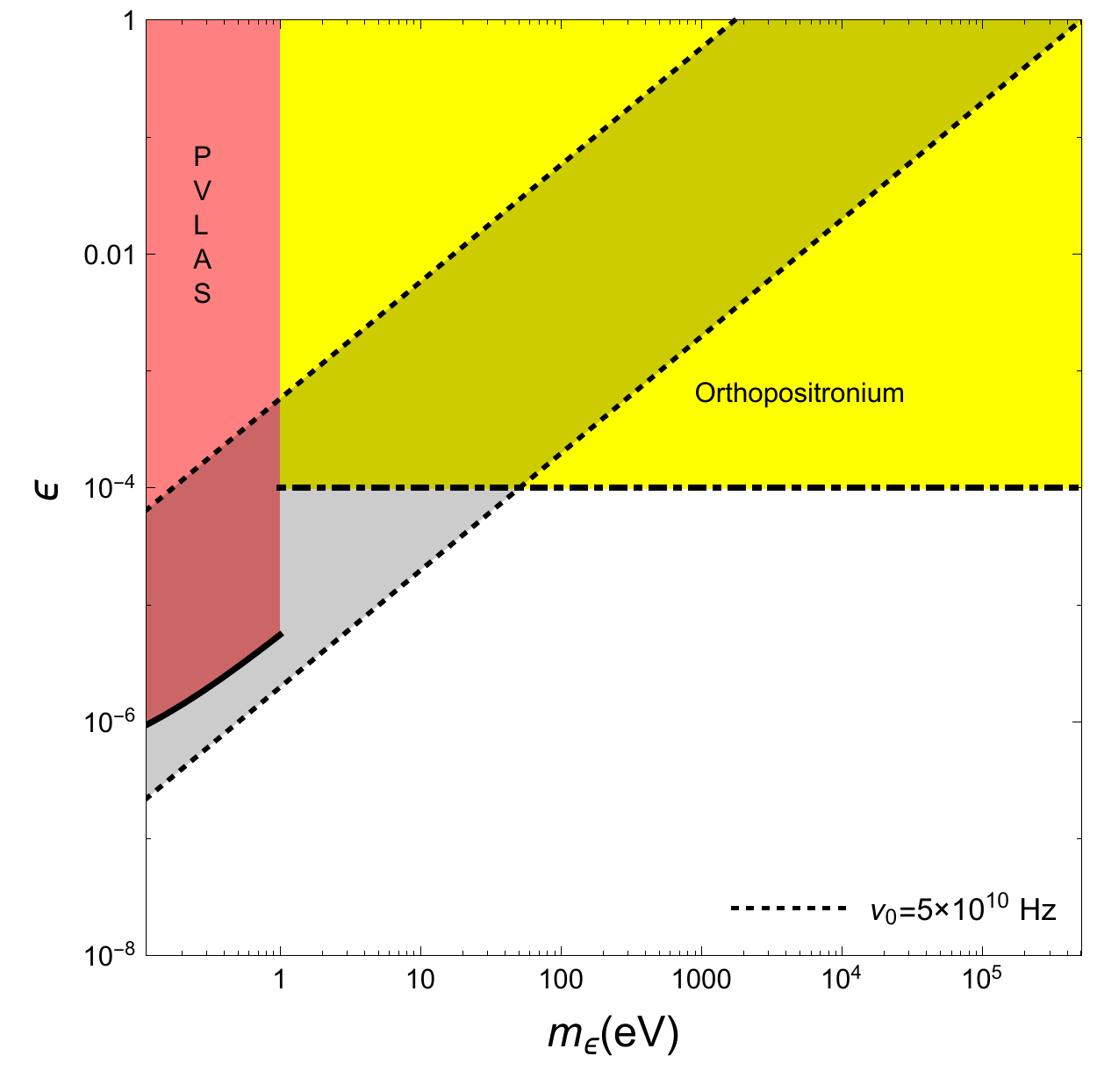}}}
\caption{In (a) the \emph{allowed} region in grey within dotted lines, in the parameter space $\epsilon$ vs. $m_\epsilon$ for magnetic field amplitude $B_{e0}=$ 1 nG and photon frequency $\nu_0=10^8$ Hz is shown. In (b) the \emph{allowed} region in grey within dotted lines for $B_{e0}=$ 1 nG and $\nu_0=5\times 10^{10}$ Hz is shown. In both plots, the \emph{exclusion} regions obtained by PVLAS \cite{DellaValle:2015xxa} (the region in pink above thick line), BFRT \cite{Gies:2006ca} (the region in green above thick dashed line) and invisible decays of Orthopositronium \cite{Mitsui:1993ha} (the region in yellow above thick dot dashed line) are also shown. Our allowed regions have been obtained by requiring that degree of circular polarization of CMB for $\Phi=\pi/2$ is smaller than $|U_i/I_i|$ and for photons that do not decay into milli-charged fermions at post decoupling epoch at given frequency $\nu_0$,  namely $m_\epsilon\geq 2.24\times 10^{-12}(\nu_0/\text{Hz})$ eV. This situation corresponds to the case when $\mathcal G_0(T_0)=0$ at post decoupling epoch where only birefringence effect does occur. The regions in white are not excluded by our model but simply they are regions where our model \emph{joint} constraints are not valid.}
\label{fig:Fig1a}
\end{figure*}

All above considerations for $\mathcal G_0(T_0)=0$ can be done more formal by defining $\sigma\equiv \epsilon(m_e/m_\epsilon)$, which numerical value essentially fixes the ratio of $\epsilon/m_\epsilon$ and is constrained from below to be $\sigma\geq 1$. In the case when condition \eqref{weak-cond} is satisfied we have that $\sigma\ll 4.38\,(\text{Hz}/\nu_0)^{1/4}(\text{G}/B_{e0})^{1/2}$ while in the case when expression \eqref{unity-cond} applies, the value of $\sigma$ is fixed. Now by using using the condition \eqref{chi-cond}, the condition of decay into milli-charged fermions before decoupling for observation frequency $\nu_0$, the conditions \eqref{landau-cond}-\eqref{critic-cond} and $\sigma\ll 4.38\,(\text{Hz}/\nu_0)^{1/4}(\text{G}/B_{e0})^{1/2}$, we get the following solutions
\begin{equation}\label{solu-1}
\begin{gathered}
4.4\times 10^{-18}\, \left(\frac{\nu_0}{\text{Hz}}\right)  \leq  \epsilon\ll 1.923\times 10^{-17} \left(\frac{\nu_0}{\text{Hz}}\right)^{3/4}\left(\frac{\text G}{B_{e0}}\right)^{1/2}\, \text{for}\quad 2.24\times 10^{-12}\left(\frac{\nu_0}{\text{Hz}}\right) \text{eV}\leq m_\epsilon \leq 5.1\times 10^5\,\epsilon\,\text{eV}\\  \text{or}\quad 
1.923\times 10^{-17} \left(\frac{\nu_0}{\text{Hz}}\right)^{3/4}\left(\frac{\text G}{B_{e0}}\right)^{1/2} \leq  \epsilon\ll \text{min}\left\{1,\, 5.95\times 10^{-28}\left(\frac{\nu_0}{\text{Hz}}\right)^2\left(\frac{\text{G}}{B_{e0}}\right)\right\}\quad \text{for} \\
 1.164\times 10^{5}\,\epsilon\,\left(\frac{\nu_0}{\text{Hz}}\right)^{1/4}\left(\frac{B_{e0}}{\text{G}}\right)^{1/2} \text{eV}\ll m_\epsilon \leq 5.1\times 10^5\,\epsilon\,\text{eV}.
\end{gathered}
\end{equation}
The above constraints on the parameters enforces that degree of circular polarization is smaller than $|U_i/I_i|$ for $\sigma \ll 4.38\,(\text{Hz}/\nu_0)^{1/4}(\text{G}/B_{e0})^{1/2}$. In the case when $\sigma$ is fixed such as in \eqref{unity-cond}, the condition that degree of circular polarization is equal to $|U_i/I_i|$ together with \eqref{chi-cond}-\eqref{critic-cond} translates into
\begin{gather}
1.923\times 10^{-17}\,\left[2\pi n+\frac{1}{2}\right]^{1/4}  \left(\frac{\nu_0}{\text{Hz}}\right)^{3/4}\,\left(\frac{\text{G}}{B_{e0}}\right)^{1/2}  \leq  \epsilon\ll \text{min}\left\{1,\, 5.95\times 10^{-28}\left(\frac{\nu_0}{\text{Hz}}\right)^2\left(\frac{\text{G}}{B_{e0}}\right)\right\}\quad \text{and}\nonumber\\
 m_\epsilon = 1.164\times 10^5\,\left[2\pi n+\frac{1}{2}\right]^{-1/4} \,\epsilon\,\left(\frac{\nu_0}{\text{Hz}}\right)^{1/4}\left(\frac{B_{e0}}{\text{G}}\right)^{1/2} \text{eV}.
\end{gather}
In Fig. \ref{fig:Fig1a}, the solutions \eqref{solu-1} for two observation frequencies $\nu_0=10^8$ Hz and $\nu_0=50$ GHz is shown. The regions in grey are those allowed by our constraints discussed above and those in pink and yellow are those excluded experimentally. The points with highest ratio of $\epsilon/m_\epsilon$ within the allowed grey regions and not excluded by experiments, give highest values of $P_C(T_0)$. The constraints used to find the allowed regions for our model are \emph{not} experimental constraints on the degree of circular polarization. Here one must pay attention by what do we mean by \emph{allowed} regions of our model. The region in white in both Figs. \ref{fig:Fig1} and \ref{fig:Fig2}, represent the regions of points $\epsilon$ and $m_\epsilon$ where our constraints \eqref{chi-cond}-\eqref{critic-cond} and $\sigma \ll 4.38\,(\text{Hz}/\nu_0)^{1/4}(\text{G}/B_{e0})^{1/2}$ are not valid. However, this does not mean that it is excluded by our model. The only excluded regions in Fig. \ref{fig:Fig1a} are those by experiments.

\begin{figure*}[htbp!]
\centering
\mbox{
\subfloat[\label{fig:Fig3}]{\includegraphics[scale=0.65]{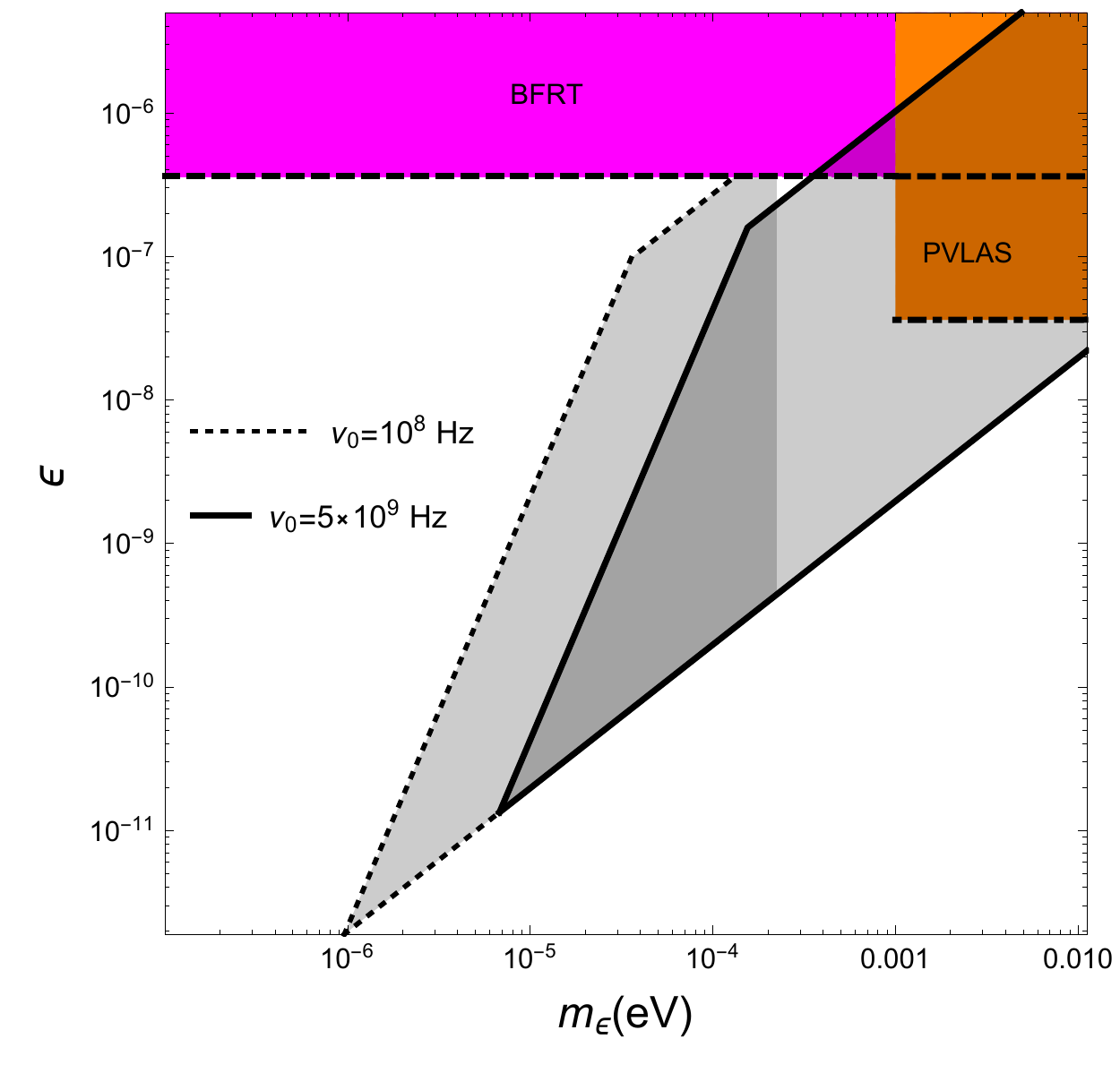}}\qquad
\subfloat[\label{fig:Fig4}]{\includegraphics[scale=0.65]{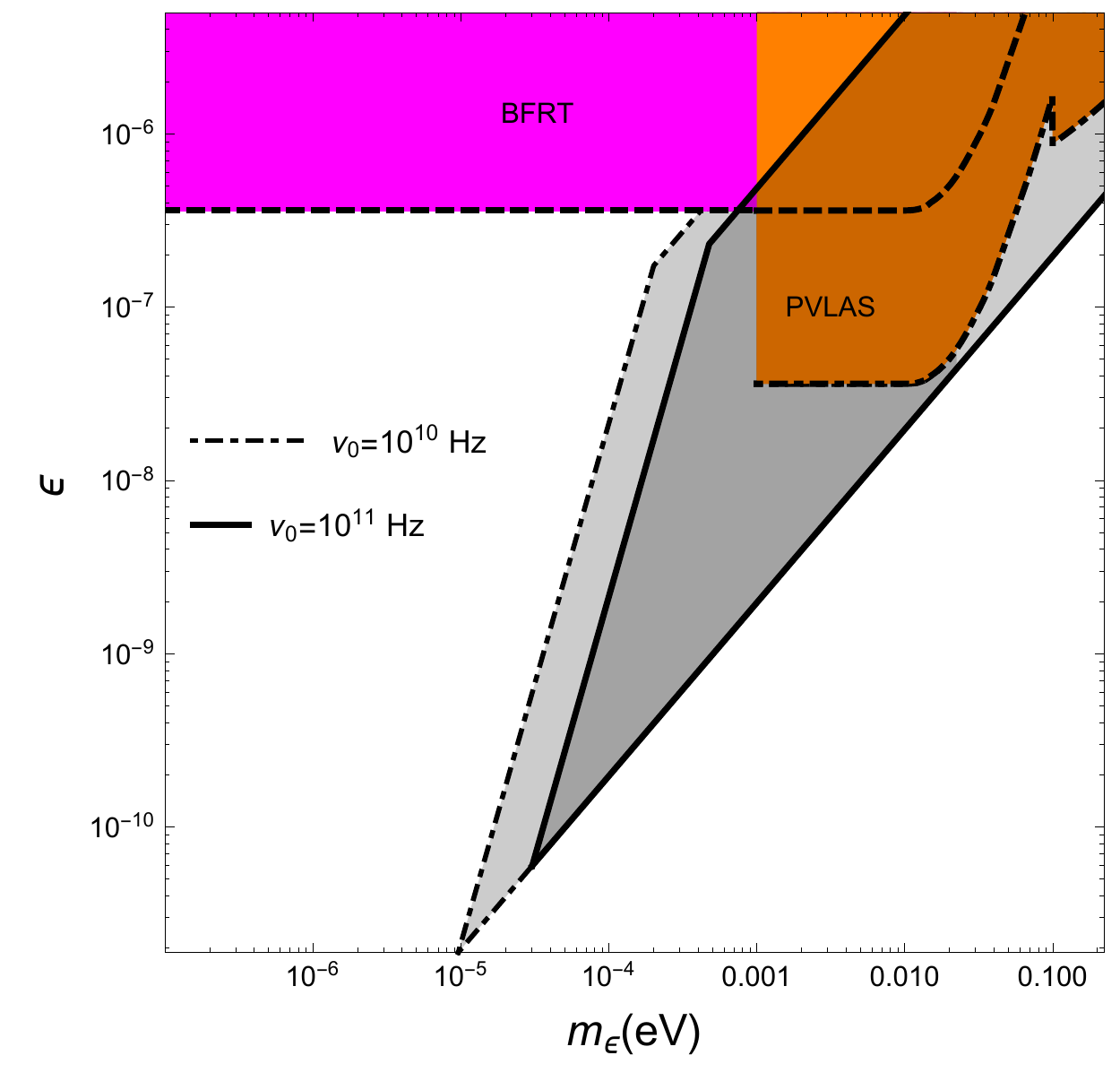}}}
\caption{In (a) \emph{allowed} regions in grey within dotted and thick lines, in the parameter space $\epsilon$ vs. $m_\epsilon$ for magnetic field amplitude $B_{e0}=$ 1 nG, $\Phi=\pi/2$, photon frequencies $\nu_0=10^8$ Hz and $\nu_0=5\times 10^9$ Hz are shown. In (b) \emph{allowed} regions in grey within dotted and thick lines for $B_{e0}=$ 1 nG, $\nu_0=10^{10}$ Hz and $\nu_0=10^{11}$ Hz are shown. In both plots, the \emph{exclusion} regions obtained by PVLAS \cite{DellaValle:2015xxa} (the region in orange above dot dashed thick line) and BFRT \cite{Gies:2006ca} (the region in magenta above dashed thick line) are also shown. Our allowed regions have been obtained from solutions \eqref{con-1} and for photons that decay into milli-charged fermions at post decoupling epoch at given frequency $\nu_0$,  namely $m_\epsilon< 2.24\times 10^{-12}(\nu_0/\text{Hz})$ eV. As in Fig. \ref{fig:Fig1a} the regions in white are not excluded by our model but simply are regions where our joint constraints are not valid.}
\label{fig:Fig2a}
\end{figure*}

\begin{figure*}[htbp!]
\centering
\mbox{
\subfloat[\label{fig:Fig5}]{\includegraphics[scale=0.65]{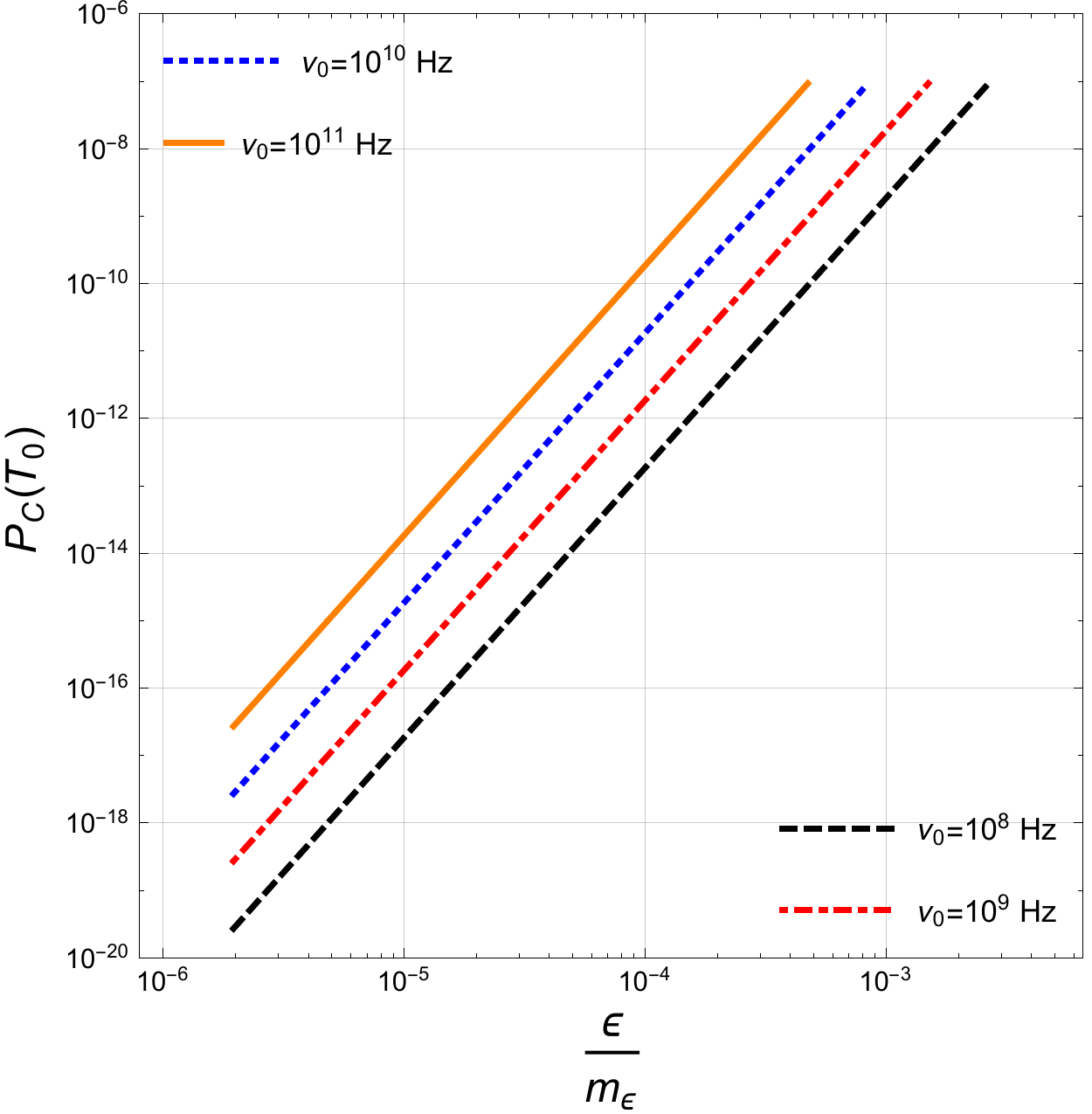}}\qquad
\subfloat[\label{fig:Fig6}]{\includegraphics[scale=0.65]{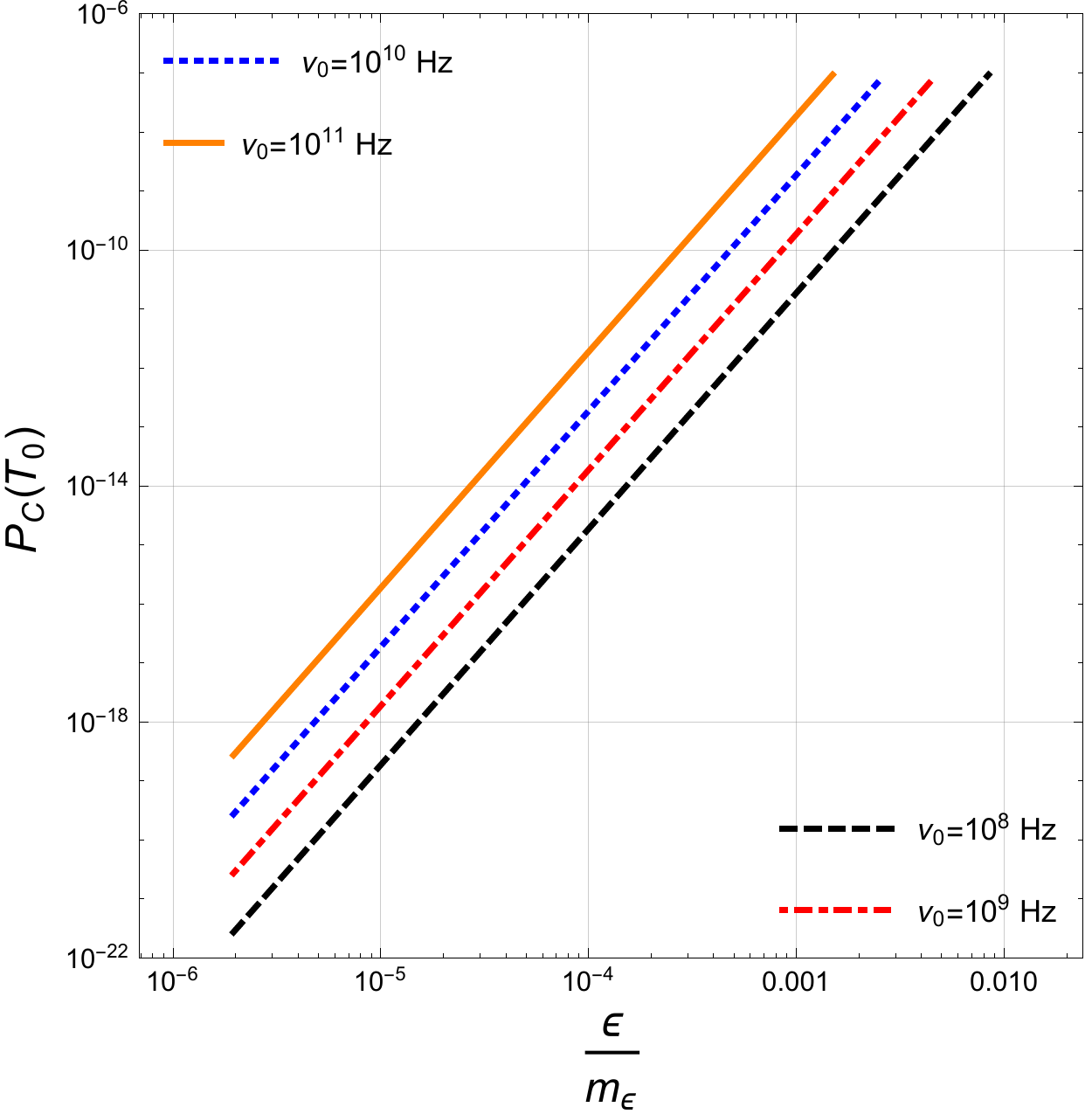}}}
\caption{Plot of the degree of circular polarization at present $P_C(T_0)$ as a function of the ratio $\epsilon/m_\epsilon$ (in units of eV$^{-1}$) given by expression \eqref{deg-circ-pol-3}. In (a) the plot for frequencies $\nu_0=\{10^8, 10^9, 10^{10}, 10^{11}\}$ Hz, magnetic field amplitude $B_{e0}=1$ nG, $|r|=0.1$ and $Q_i/I_i=10^{-6}$ is shown. In (b) the plot for the same values of frequencies and $|r|$ as in (a) with $B_{e0}=0.1$ nG is shown. }
\label{fig:Fig3a}
\end{figure*}

In the case when $|\mathcal G_0(T_0)|\ll 1$, the situation is slightly different since now decay into milli-charged fermions can occur at post decoupling epoch, namely their mass must satisfy $m_\epsilon< 2.24\times 10^{-12}(\nu_0/\text{Hz})$ eV for observation frequency $\nu_0$. Either $\mathcal G_0(T_0)$ is small or large depends exclusively on the value of $\chi(T_i)$ in expression \eqref{G-0-2}. As already mentioned above, if for example $\chi(T_i)\leq 0.01$, the value of $\mathcal G_0(T)$ is extremely small. In such case, the condition $\chi(T_i)\leq 0.01$ is satisfied when 
\begin{equation}\label{con-0}
\epsilon\left(\frac{m_e}{m_\epsilon}\right)^3\leq 2.81\times 10^{22}\left(\frac{\text{Hz}}{\nu_0}\right)\,\left(\frac{\text{G}}{B_{e0}}\right).
\end{equation}
The difference between the case when $\mathcal G_0(T_0)=0$ and $\mathcal G_0(T_0)\neq 0$, relies on the fact that in the former case $m_\epsilon\geq 2.24\times 10^{-12}(\nu_0/\text{Hz})$ eV while in the latter case $m_\epsilon< 2.24\times 10^{-12}(\nu_0/\text{Hz})$ eV. Therefore, in the case when  $1\leq \sigma\ll 4.38\,(\text{Hz}/\nu_0)^{1/4}(\text{G}/B_{e0})^{1/2}$, the condition that the degree of circular polarization is smaller than $|U_i/I_i|$ together with conditions \eqref{con-0}, \eqref{landau-cond}-\eqref{critic-cond} and $m_\epsilon< 2.24\times 10^{-12}(\nu_0/\text{Hz})$ eV, is satisfied when
\begin{equation}\label{con-1}
\begin{gathered}
5.96\times 10^{-12} \left(\frac{B_{e0}}{\text{G}}\right)^{1/2}\,\left(\frac{\nu_0}{\text{Hz}}\right)^{1/2} \leq \epsilon <  4.4\times 10^{-18}\, \left(\frac{\nu_0}{\text{Hz}}\right)\quad \text{for}\quad  R \leq \left(\frac{m_\epsilon}{\text{eV}}\right) \leq 5.1\times 10^5\, \epsilon\quad \text{or}\quad  \\ 
4.4\times 10^{-18}\, \left(\frac{\nu_0}{\text{Hz}}\right)\leq \epsilon< 5.46\times 10^{-11}\,  \left(\frac{B_{e0}}{\text{G}}\right)^{-1/4}\,\left(\frac{\nu_0}{\text{Hz}}\right)^{1/8}\quad \text{for}\quad   R \leq \left(\frac{m_\epsilon}{\text{eV}}\right)< 2.24\times 10^{-12}\left(\frac{\nu_0}{\text{Hz}}\right)\quad \text{or} \\
5.46\times 10^{-11}\,  \left(\frac{B_{e0}}{\text{G}}\right)^{-1/4}\,\left(\frac{\nu_0}{\text{Hz}}\right)^{1/8} \leq \epsilon< 1.92\times 10^{-17}\,  \left(\frac{B_{e0}}{\text{G}}\right)^{-1/2}\,\left(\frac{\nu_0}{\text{Hz}}\right)^{3/4}\quad \text{for}\quad \\ 1.164\times 10^{5}\,\epsilon\, \left(\frac{B_{e0}}{\text{G}}\right)^{1/2}\,\left(\frac{\nu_0}{\text{Hz}}\right)^{1/4} < \left(\frac{m_\epsilon}{\text{eV}}\right)< 2.24\times 10^{-12}\left(\frac{\nu_0}{\text{Hz}}\right),
\end{gathered}
\end{equation}
where $R$ is the root of the equation $2.12\times 10^5 y^3-(\nu_0/\text{Hz})(B_{e0}/\text{G})\,\epsilon=0$ with $y$ being the dependent variable and $\nu_0, B_{e0}, \epsilon$ being the independent variables. In Fig. \ref{fig:Fig2a} the allowed regions of parameters given by solutions \eqref{con-1} are shown. In Fig. \ref{fig:Fig3a}, plots of degree of circular polarization at present as a function of $\epsilon/m_e$, for different values of the frequency $\nu_0$ and magnetic field strength $B_{e0}$ are shown. One can easily check from Figs. \ref{fig:Fig2a} and \ref{fig:Fig3a} that values of $\epsilon$ and $m_\epsilon$, that form the ratio $\epsilon/m_\epsilon$ plotted in Fig. \ref{fig:Fig3a} corresponding to a given frequency $\nu_0$, are within the allowed regions of Fig. \ref{fig:Fig2a} for given frequency $\nu_0$.

In the case when $\sigma$ is fixed by \eqref{unity-cond}, the condition that the degree of circular polarization is equal to $|U_i/I_i|$ together with conditions \eqref{con-0}, \eqref{landau-cond}-\eqref{critic-cond} and $m_\epsilon< 2.24\times 10^{-12}(\nu_0/\text{Hz})$ eV, is satisfied when
\begin{gather}\label{con-2}
5.46\times 10^{-11}\,\left[2\pi n+\frac{1}{2}\right]^{3/8}  \left(\frac{B_{e0}}{\text{G}}\right)^{-1/4}\,\left(\frac{\nu_0}{\text{Hz}}\right)^{1/8}\leq \epsilon < 1.923\times 10^{-17}\,\left[2\pi n+\frac{1}{2}\right]^{1/4} \left(\frac{\nu_0}{\text{Hz}}\right)^{3/4}\,\left(\frac{\text{G}}{B_{e0}}\right)^{1/2}\nonumber\\
 \text{for}\quad m_\epsilon = 1.164\times 10^{5}\,\left[2\pi n+\frac{1}{2}\right]^{-1/4}\,\epsilon\,\left(\frac{\nu_0}{\text{Hz}}\right)^{1/4}\left(\frac{B_{e0}}{\text{G}}\right)^{1/2} \text{eV}.
\end{gather}

In addition to generation of circular and linear polarization\footnote{It is easy to check from solutions \eqref{sol-no-far} that there is generation of linear polarization even in the case when $Q_i=U_i=V_i=0$.}, there is also rotation of the polarization plane of the CMB. In general, if light is initially polarized, the rotation angle of polarization plane is given by $\tan[2\psi]=U/Q$. In the case when the CMB is initially only linearly polarized, from solutions \eqref{sol-no-far}, we get the following general expression
\begin{equation}\label{rot-pol-plane}
\tan[2\psi(T)]=\frac{U(T)}{Q(T)}=\frac{2\,\cos[\tilde{\mathcal G}(T)]U_i}{(\exp[-\mathcal G_0(T)]-\exp[\mathcal G_0(T)]) I_i+(\exp[-\mathcal G_0(T)]+\exp[\mathcal G_0(T)]) Q_i}.
\end{equation}
In the case when $|\mathcal G_0(T)|\ll 1$ or $\mathcal G_0(T)=0$, we would get
\begin{equation}\label{rot-pol-plane-1}
\tan[2\psi(T)]\simeq \tan[2\psi(T_i)]\cos[\tilde{\mathcal G}(T)].
\end{equation}
We may write $\psi(T)=\psi(T_i)+\delta\psi(T)$ where $|\delta\psi(T)|\ll 1$ and consequently use $\tan[2\psi(T_i)+2\,\delta\psi(T)]=  r+2\,\delta\psi(T)(1+r^2)+O(\delta\psi^2)$ where we have defined\footnote{The parameter $r$ in principle can have either sign and its value is not known. In this work we assume that $|r|\leq 1$.} $r\equiv U_i/Q_i$. Consequently from expression \eqref{rot-pol-plane-1} we get
\begin{equation}\label{rot-angle}
\delta\psi(T)\simeq \frac{- r}{2(1+r^2)}\,\left(1-\cos[\tilde{\mathcal G}(T)]\right)\simeq\frac{-r}{4(1+r^2)}\tilde{\mathcal G}^2(T),
\end{equation}
where we used the fact that one would expect very small rotation angle $\delta\psi$ and consequently $\tilde{\mathcal G}(T)\ll 1$. Using the expression for $\tilde{\mathcal{G}}(T_0)$, we get
\begin{equation}\label{rot-angle-1}
\epsilon\left(\frac{m_e}{m_\epsilon}\right)\simeq 4.38\,\left[\frac{-4\,\delta\psi(T_0)(1+r^2)}{r}\right]^{1/8}\left(\frac{\text{Hz}}{\nu_0}\right)^{1/4}\left(\frac{\text{G}}{B_{e0}}\right)^{1/2}.
\end{equation}
In order to enforce positivity in expression \eqref{rot-angle-1}, if the rotation angle is negative $\delta\psi<0$ we have that $U_i$ must have the same sign as $Q_i$. If $\delta\psi>0$ we must have that $U_i$ and $Q_i$ have opposite signs.

Current limits on the rotation angle of the CMB polarization plane have been found by different experiments and for a review see Ref. \cite{Galaverni:2014gca}. Here we consider for simplicity only the result found by WMAP9 \cite{Hinshaw:2012aka} in the case of uniform rotation across the sky\footnote{The limit found by WMAP9 \cite{Hinshaw:2012aka} is very close to the current limit found by Planck collaboration \cite{Aghanim:2016fhp} under the hypothesis of uniform rotation across the sky. However, the limit presented by WMAP  \cite{Hinshaw:2012aka} has been found for $\nu_0=$ 53 GHz while the limit presented by Planck collaboration \cite{Aghanim:2016fhp} apparently seems frequency independent.}, $\delta\psi(T_0)\simeq -0.36^\circ=-6.28\times 10^{-3}$ (rad) at the observation frequency $\nu_0=53$ GHz. Expression \eqref{rot-angle-1} has been derived in the case when light propagates perpendicular to the external magnetic field, namely is has been derived for a specific direction of propagation. If one observes the CMB in another direction, one would get another expression for the rotation angle. This is due to the fact that rotation angle $\delta\psi(T)$ depends on $\Phi$ and consequently in our model of milli-charged fermion vacuum polarization it is not uniform across the sky. However, for an order of magnitude estimate we may use current WMAP9 limit on $\delta\psi(T_0)$ in expression \eqref{rot-angle-1}. Therefore we get
\begin{equation}\label{rot-angle-2}
\epsilon\left(\frac{m_e}{m_\epsilon}\right)\simeq 5.76\times 10^{-3}\,\left[\frac{1+r^2}{r}\right]^{1/8}\left(\frac{\text{G}}{B_{e0}}\right)^{1/2}.
\end{equation}
Let us consider for example the case when $B_{e0}=1$ nG and $r\simeq 1$, which would give $\epsilon(m_e/m_\epsilon)\simeq 198.6$. If we consider $r=0.1$ with the same value of $B_{e0}=1$ nG  we would get $\epsilon(m_e/m_\epsilon)\simeq 243.2$.

The linear relation between milli-charged fermion mass $m_\epsilon$ and $\epsilon$ given in  \eqref{rot-angle-1} can be used to express $P_C(T_0)$ in terms of the rotation angle $\delta\psi(T_0)$ and $r$. Indeed, if $|\delta\psi(T_0)|< |r|/4(1+r^2)$, we have that the right hand side of \eqref{rot-angle-1} is smaller than the right hand side of expression \eqref{weak-cond}. Then by using expression \eqref{rot-angle-1} into expression \eqref{deg-circ-pol-3}, we get the following relation for $P_C(T_0)$
\begin{equation}\label{rel-PC-psi}
P_C(T_0)\simeq \left[-4\,\delta\psi(T_0) r (1+r^2)\right]^{1/2}\left|\frac{Q_i}{I_i}\right|.
\end{equation}
It is very interesting to note from expression \eqref{rel-PC-psi} that $P_C(T_0)$ does not explicitly depend neither on $B_{e0}$ nor on $\nu_0$. For example if we use $r=1$ and the value of $\delta\psi(T_0)=-6.28\times 10^{-3}$ obtained by WMAP9 \cite{Hinshaw:2012aka}, we would get $P_C(T_0)\simeq 0.22|Q_i/I_i|\simeq 2.2\times 10^{-7}$ for $|Q_i/I_i|\simeq 10^{-6}$. In the case of $r=0.1$, we get $P_C(T_0)\simeq 0.05|Q_i/I_i|\simeq 5\times 10^{-8}$ for $|Q_i/I_i|\simeq 10^{-6}$. The expression \eqref{rot-angle-1} has been derived under the hypothesis $|\mathcal G_0(T)|\ll 1$ or $\mathcal G_0(T)=0$. With the value of $\sigma$ fixed to
\begin{equation}\nonumber
\sigma=4.38\,\left[\frac{-4\,\delta\psi(T_0)(1+r^2)}{r}\right]^{1/8}\left(\frac{\text{Hz}}{\nu_0}\right)^{1/4}\left(\frac{\text{G}}{B_{e0}}\right)^{1/2},
\end{equation}
the condition that the degree of circular polarization is less than $|U_i/I_i|$ together with conditions \eqref{con-0}, \eqref{critic-cond} and $m_\epsilon< 2.24\times 10^{-12}(\nu_0/\text{Hz})$ eV, are satisfied when
\begin{gather}\label{con-3}
5.46\times 10^{-11}\,\left[\frac{-4\,\delta\psi(T_0)(1+r^2)}{r}\right]^{3/16} \left(\frac{B_{e0}}{\text{G}}\right)^{-1/4}\,\left(\frac{\nu_0}{\text{Hz}}\right)^{1/8}\leq \epsilon < 1.923\times 10^{-17}\,\left[\frac{-4\,\delta\psi(T_0)(1+r^2)}{r}\right]^{1/8} \nonumber\\ \times \left(\frac{\nu_0}{\text{Hz}}\right)^{3/4}\left(\frac{\text{G}}{B_{e0}}\right)^{1/2}\quad
 \text{for}\quad m_\epsilon = 1.164\times 10^{5}\,\left[\frac{-4\,\delta\psi(T_0)(1+r^2)}{r}\right]^{-1/8}\,\epsilon\,\left(\frac{\nu_0}{\text{Hz}}\right)^{1/4}\left(\frac{B_{e0}}{\text{G}}\right)^{1/2} \text{eV}.
\end{gather}
In the case when $\mathcal G_0(T_0)=0$ or equivalently $m_\epsilon\geq 2.24\times 10^{-12}(\nu_0/\text{Hz})$ eV, the conditions that the degree of circular polarization is less than $|U_i/I_i|$ together with conditions \eqref{chi-cond}-\eqref{critic-cond} are satisfied for 
\begin{gather}\label{con-4}
1.923\times 10^{-17}\,\left[\frac{-4\,\delta\psi(T_0)(1+r^2)}{r}\right]^{1/8} \left(\frac{\nu_0}{\text{Hz}}\right)^{3/4}\left(\frac{\text{G}}{B_{e0}}\right)^{1/2}\leq \epsilon\ll \text{min}\left\{1,\, 5.95\times 10^{-28}\left(\frac{\nu_0}{\text{Hz}}\right)^2\left(\frac{\text{G}}{B_{e0}}\right)\right\}\nonumber\\
 \text{for}\quad m_\epsilon = 1.164\times 10^{5}\,\left[\frac{-4\,\delta\psi(T_0)(1+r^2)}{r}\right]^{-1/8}\,\epsilon\,\left(\frac{\nu_0}{\text{Hz}}\right)^{1/4}\left(\frac{B_{e0}}{\text{G}}\right)^{1/2} \text{eV}.
\end{gather}

\subsection{Generation of polarization in case of $\chi\gg 1$.}
\label{sec:subsection-1}

In the case when $\chi(T)\gg 1$, we get
\begin{equation}\label{con-5}
\epsilon\left(\frac{m_e}{m_\epsilon}\right)^3\gg 3.65\times 10^{33}\left(\frac{\text{Hz}}{\nu_0}\right)\,\left(\frac{\text{G}}{B_{e0}}\right)\left(\frac{T_0}{T}\right)^3,
\end{equation}
and it is fulfilled for all $T_0\leq T\leq T_i$ only when $T=T_0$. Consequently, we can write conditions \eqref{con-5} and \eqref{critic-cond} as constraints on the mass $m_\epsilon$
\begin{equation}\label{con-6}
83.42\, \epsilon^{1/2}\left(\frac{B_{e0}}{\text{G}}\right)^{1/2}|\sin(\Phi)|^{1/2}\quad  \ll \left(\frac{m_\epsilon}{\text{eV}}\right)\ll 3.31\times 10^{-6}\epsilon^{1/3}\left(\frac{\nu_0}{\text{Hz}}\right)^{1/3}\,\left(\frac{B_{e0}}{\text{G}}\right)^{1/3}.
\end{equation}
On the other hand, condition \eqref{con-6} is satisfied for values of $\epsilon$ satisfying the constriant $\epsilon\ll 3.9\times 10^{-45}|\sin(\Phi)|^{-3}(\nu_0/\text{Hz})^2(\text{G}/B_{e0})$. If we take for example $B_{e0}=1$ nG and $\nu_0=100$ GHz, we would get $\epsilon\ll 3.9\times 10^{-14}|\sin(\Phi)|^{-3}$. It is quite straightforward to check that for such small values of $\epsilon$, we get also very small values for $m_\epsilon$ from expression \eqref{con-6}. For these small values of the parameters $\epsilon$ and $m_\epsilon$, all quantities of interest such as $P_C(T_0)$ and/or $\delta\psi(T_0)$, are extremely small with respect to the case when $\chi(T)\ll 1$ and of not practical interest. Indeed, as we explicitly have checked, the quantities of interest are proportional to fractional powers of $\epsilon$ only and do not explicitly depend on $m_\epsilon$. This situation is opposite to the case of $\chi(T)\ll 1$, where quantities of interest are proportional to some powers of the ratio ($\epsilon/m_\epsilon$).

\section{Solutions of equations of motions in case of $\Phi\neq \pi/2$.}
\label{sec:5}

In this section we consider the case when the angle of observation is $\Phi\neq \pi/2$. In this case in addition to the effect caused milli-charged fermion vacuum polarization, there is also present the Faraday effect since the external magnetic field has a longitudinal component along the observation direction. Now in order to treat generation of CMB polarization, we must solve Eq. \eqref{eq-stokes-vec} by using perturbation theory as explained in Sec. \ref{sec:2}. Now we are faced with the problem of how to split the matrix $B(T)$ in order to use perturbation theory. In addition we must consider the fact that our results will depend on the angle $\Phi$. 

In matrix $B(T)$ enter three terms $\Delta M_0,  2M_F$ and $\Delta\tilde M$ which relative magnitude will depend on $\epsilon$ and $m_\epsilon$ for fixed $\nu_0$ and $B_{e0}$. Consider the case when the term corresponding to the Faraday effect is much bigger than $\Delta M_0$ and $\Delta\tilde M$. In this case in $B_0(T)$ enter only the term due to the Faraday effect while in $\lambda B_1(T)$ enter the remaining terms, namely $\Delta M_0$ and $\Delta\tilde M$. Since the latter two terms are assumed to be smaller than the Faraday term and because we are interested in exploring a vast parameter space of $m_\epsilon$ and $\epsilon$, it is necessary to look for solution of the Stokes vector up to second order in perturbation theory. So, by using matrix equations \eqref{perturbative-sol}, we get the following solutions for the Stokes parameters up to second order in perturbation theory
\begin{align}
\left(\frac{T_i}{T}\right)^2\exp[\mathcal G_1(T)]I(T) &= \left[1+\int_T^{T_i}dT^\prime\,G_0(T^\prime)\cos[\mathcal M_F(T^\prime)]\, \int_{T^\prime}^{T_i}dT^{\prime\prime}\,G_0(T^{\prime\prime})\cos[\mathcal M_F(T^{\prime\prime})]+ \int_T^{T_i}dT^\prime\,G_0(T^\prime)\right.\nonumber\\ & \left. \times \sin[\mathcal M_F(T^\prime)]\,\int_{T^\prime}^{T_i}dT^{\prime\prime}\,G_0(T^{\prime\prime})\sin[\mathcal M_F(T^{\prime\prime})]\right]\,I_i-\left(\int_T^{T_i}dT^\prime\,G_0(T^\prime)\cos[\mathcal M_F(T^\prime)]\right)\, Q_i\nonumber\\ & + \left(\int_T^{T_i}dT^\prime\,G_0(T^\prime)\sin[\mathcal M_F(T^\prime)]\right)\, U_i,\nonumber\\
\left(\frac{T_i}{T}\right)^2\exp[\mathcal G_1(T)]Q(T) &= -\left[\cos[\mathcal M_F(T)]\int_{T}^{T_i}dT^{\prime}\,G_0(T^{\prime})\cos[\mathcal M_F(T^{\prime})]+ \sin[\mathcal M_F(T^\prime)]\int_{T}^{T_i}dT^{\prime}\,G_0(T^{\prime})\times\right.\nonumber\\ & \left. \sin[\mathcal M_F(T^{\prime})]\right]\,I_i+\left[\cos[\mathcal M_F(T)]+\cos[\mathcal M_F(T)]\left(\int_T^{T_i}dT^\prime\,G_0(T^\prime)\cos[\mathcal M_F(T^\prime)]\right.\right.\nonumber\\ & \left.\left. \times \int_{T^\prime}^{T_i}dT^{\prime\prime}\,G_0(T^{\prime\prime})\cos[\mathcal M_F(T^{\prime\prime})]- \int_T^{T_i}dT^\prime\,\tilde G(T^\prime) \sin[\mathcal M_F(T^\prime)]\,\int_{T^\prime}^{T_i}dT^{\prime\prime}\,\tilde G(T^{\prime\prime})\right.\right.\nonumber\\ & \left.\left. \times\sin[\mathcal M_F(T^{\prime\prime})]\right)+\sin[\mathcal M_F(T)]\left(\int_T^{T_i}dT^\prime\,G_0(T^\prime)\sin[\mathcal M_F(T^\prime)] \right.\right.\nonumber\\ & \left.\left. \times \int_{T^\prime}^{T_i}dT^{\prime\prime}\,G_0(T^{\prime\prime})\cos[\mathcal M_F(T^{\prime\prime})]+ \int_T^{T_i}dT^\prime\,\tilde G(T^\prime) \cos[\mathcal M_F(T^\prime)]\,\int_{T^\prime}^{T_i}dT^{\prime\prime}\,\tilde G(T^{\prime\prime})\right.\right.\nonumber\\ & \left.\left. \times\sin[\mathcal M_F(T^{\prime\prime})]\right)\right]Q_i-\left[\sin[\mathcal M_F(T)]-\sin[\mathcal M_F(T)]\left(\int_T^{T_i}dT^\prime\,\tilde G(T^\prime)\cos[\mathcal M_F(T^\prime)] \right.\right.\nonumber\\ & \left.\left. \times \int_{T^\prime}^{T_i}dT^{\prime\prime}\,\tilde G(T^{\prime\prime})\cos[\mathcal M_F(T^{\prime\prime})]- \int_T^{T_i}dT^\prime\,G_0(T^\prime) \sin[\mathcal M_F(T^\prime)]\,\int_{T^\prime}^{T_i}dT^{\prime\prime}\,G_0(T^{\prime\prime})\right.\right.\nonumber\\ & \left.\left. \times\sin[\mathcal M_F(T^{\prime\prime})]\right)-\cos[\mathcal M_F(T)]\left(\int_T^{T_i}dT^\prime\,G_0(T^\prime)\cos[\mathcal M_F(T^\prime)] \right.\right.\nonumber\\ & \left.\left. \times \int_{T^\prime}^{T_i}dT^{\prime\prime}\,G_0(T^{\prime\prime})\sin[\mathcal M_F(T^{\prime\prime})]+ \int_T^{T_i}dT^\prime\,\tilde G(T^\prime) \sin[\mathcal M_F(T^\prime)]\,\int_{T^\prime}^{T_i}dT^{\prime\prime}\,\tilde G(T^{\prime\prime})\right.\right.\nonumber\\ & \left.\left. \times\cos[\mathcal M_F(T^{\prime\prime})]\right)\right]U_i+\left[\cos[\mathcal M_F(T)]\,\int_T^{T_i}dT^\prime\,\tilde G(T^\prime)\sin[\mathcal M_F(T^\prime)]-\right.\nonumber\\ & \left. \sin[\mathcal M_F(T)]\,\int_T^{T_i}dT^\prime\,\tilde G(T^\prime)\cos[\mathcal M_F(T^\prime)]\right]V_i,\nonumber
\end{align}

\begin{align}\label{pert-sol-par}
\left(\frac{T_i}{T}\right)^2\exp[\mathcal G_1(T)]U(T) &= \left[\cos[\mathcal M_F(T)]\int_T^{T_i}dT^\prime\,G_0(T^\prime)\sin[\mathcal M_F(T^\prime)]- \sin[\mathcal M_F(T)]\int_T^{T_i}dT^\prime\,G_0(T^\prime)\cos[\mathcal M_F(T^\prime)]\right]I_i\nonumber\\ & +
\left[\sin[\mathcal M_F(T)]-\sin[\mathcal M_F(T)]\left(\int_T^{T_i}dT^\prime\,\tilde G(T^\prime)\sin[\mathcal M_F(T^\prime)]  \int_{T^\prime}^{T_i}dT^{\prime\prime}\,\tilde G(T^{\prime\prime})\sin[\mathcal M_F(T^{\prime\prime})]\right.\right.\nonumber\\ & \left.\left. - \int_T^{T_i}dT^\prime\,G_0(T^\prime) \cos[\mathcal M_F(T^\prime)]\,\int_{T^\prime}^{T_i}dT^{\prime\prime}\,G_0(T^{\prime\prime})\cos[\mathcal M_F(T^{\prime\prime})]\right)-\right.\nonumber\\ & \left.\cos[\mathcal M_F(T)]\left(\int_T^{T_i}dT^\prime\,G_0(T^\prime)\sin[\mathcal M_F(T^\prime)]  \int_{T^\prime}^{T_i}dT^{\prime\prime}\,G_0(T^{\prime\prime})\cos[\mathcal M_F(T^{\prime\prime})]+ \right.\right.\nonumber\\ & \left.\left. \int_T^{T_i}dT^\prime\,\tilde G(T^\prime) \cos[\mathcal M_F(T^\prime)]\,\int_{T^\prime}^{T_i}dT^{\prime\prime}\,\tilde G(T^{\prime\prime})\sin[\mathcal M_F(T^{\prime\prime})]\right)\right]Q_i+ \nonumber\\ &
\left[\cos[\mathcal M_F(T)]-\cos[\mathcal M_F(T)]\left(\int_T^{T_i}dT^\prime\,\tilde G(T^\prime)\cos[\mathcal M_F(T^\prime)]  \int_{T^\prime}^{T_i}dT^{\prime\prime}\,\tilde G(T^{\prime\prime})\cos[\mathcal M_F(T^{\prime\prime})]\right.\right.\nonumber\\ & \left.\left. - \int_T^{T_i}dT^\prime\,G_0(T^\prime) \sin[\mathcal M_F(T^\prime)]\,\int_{T^\prime}^{T_i}dT^{\prime\prime}\,G_0(T^{\prime\prime})\sin[\mathcal M_F(T^{\prime\prime})]\right)-\right.\nonumber\\ & \left.\sin[\mathcal M_F(T)]\left(\int_T^{T_i}dT^\prime\,G_0(T^\prime)\cos[\mathcal M_F(T^\prime)]  \int_{T^\prime}^{T_i}dT^{\prime\prime}\,G_0(T^{\prime\prime})\sin[\mathcal M_F(T^{\prime\prime})]+ \right.\right.\nonumber\\ & \left.\left. \int_T^{T_i}dT^\prime\,\tilde G(T^\prime) \sin[\mathcal M_F(T^\prime)]\,\int_{T^\prime}^{T_i}dT^{\prime\prime}\,\tilde G(T^{\prime\prime})\cos[\mathcal M_F(T^{\prime\prime})]\right)\right]U_i+\nonumber\\ &
\left[\cos[\mathcal M_F(T)]\int_T^{T_i}dT^\prime\,\tilde G(T^\prime)\cos[\mathcal M_F(T^\prime)]+ \sin[\mathcal M_F(T)]\int_T^{T_i}dT^\prime\,\tilde G(T^\prime)\sin[\mathcal M_F(T^\prime)]\right] V_i,\nonumber\\
\left(\frac{T_i}{T}\right)^2\exp[\mathcal G_1(T)]V(T) &= -\left(\int_{T}^{T_i}dT^{\prime}\,\tilde G(T^{\prime})\sin[\mathcal M_F(T^{\prime})]\right)Q_i - \left(\int_{T}^{T_i}dT^{\prime}\,\tilde G(T^{\prime})\cos[\mathcal M_F(T^{\prime})]\right)U_i \nonumber\\ & + \left[1-\int_T^{T_i}dT^\prime\,\tilde G(T^\prime)\cos[\mathcal M_F(T^\prime)]\, \int_{T^\prime}^{T_i}dT^{\prime\prime}\,\tilde G(T^{\prime\prime})\cos[\mathcal M_F(T^{\prime\prime})]- \int_T^{T_i}dT^\prime\,\tilde G(T^\prime)\right.\nonumber\\ & \left. \times \sin[\mathcal M_F(T^\prime)]\,\int_{T^\prime}^{T_i}dT^{\prime\prime}\,\tilde G(T^{\prime\prime})\sin[\mathcal M_F(T^{\prime\prime})]\right]\,V_i.
\end{align}
The expressions for $\mathcal M_F$ and $M_F$ are respectively given by \cite{Ejlli:2016avx}\begin{equation}\nonumber
\mathcal M_F(T)=\int_T^{T_i}dT^\prime\, \frac{2M_F(T^\prime)}{H(T^\prime)T^\prime}\quad \text{with}\quad M_F=\frac{|\Pi^{12}|}{2\omega}=\frac{\omega_\text{pl}^2\,\omega_c\cos(\Phi)}{2(\omega^2-\omega_c^2)},
\end{equation}
where $\omega_\text{pl}$ is the plasma frequency and in the case of CMB we essentially have $\omega\gg \omega_c$. Here we also used the notation $\tilde G(T)=\Delta\tilde M(T)/HT$ and $ G_0(T)= \Delta M_0(T)/HT$. We may note from solutions \eqref{pert-sol-par} that in the case when $G_0(T)=0$ or absence of photon decay into milli-charged fermions, we have essentially that photon intensity changes only due to universe expansion, namely $(T_i/T)^2 I=I_i$. Moreover, in case when the CMB is initially unpolarized, we have $U(T)\neq 0$ and $Q(T)\neq 0$. Therefore for $\Phi\neq \pi/2$ we have $U(T)\neq$0, which is opposite to the case when $\Phi=\pi/2$ where $U(T)=0$ and $Q(T)\neq 0$. It is also interesting to note that even in the case when $\Phi\neq\pi/2$ and for initially unpolarized CMB, we still have $V(T)=0$ from \eqref{pert-sol-par} to second order in perturbation theory.

\subsection{Generation of polarization in the case $\Phi\neq\pi/2$ and $\chi\ll 1$.}

In this section we consider generation of CMB polarization in the case when $\Phi\neq\pi/2$ and when $\chi(T)\ll 1$. As we may observe from solutions \eqref{pert-sol-par}, the CMB acquires elliptic polarization in the case when is initially polarized. In the case of $\Phi\neq\pi/2$ and $\chi(T)\ll 1$, in addition to conditions \eqref{chi-cond}-\eqref{critic-cond}, there also two additional conditions coming from the fact that $|2M_F|\gg |\Delta M_0|, |\Delta\tilde M|$. The condition $|2 M_F(T)|\gg |\Delta M_0(T)|$ is satisfied at post decoupling epoch for all $T_0\leq T\leq T_i$, if
\begin{equation}\label{far-big-pair}
\epsilon \ll 23.84 \exp[4/(3\chi)] \left(\frac{\text{Hz}}{\nu_0}\right)^{2/3}\text{min}\left\{X_e^{1/3}(T)\left(\frac{T}{T_0}\right)^{1/3}\right\}\left(\frac{m_\epsilon}{m_e}\right)^{1/3}\left|\tan(\Phi)\right|^{-1/3}, \end{equation}
while the condition $|2M_F(T)|\gg |\Delta\tilde M(T)|$ is satisfied for all $T_0\leq T\leq T_i$, if\begin{equation}\label{far-big-qed}
\epsilon \ll 5.73\times 10^{9} \left(\frac{\text{Hz}}{\nu_0}\right)^{3/4}\left(\frac{m_\epsilon}{m_e}\right)\left(\frac{\text G}{B_{e0}}\right)^{1/4}\,\text{min}\left\{X_e^{1/4}(T)\left(\frac{T}{T_0}\right)^{-1/2}\right\}\left|\frac{\cos(\Phi)}{\sin^2(\Phi)}\right|^{1/4}, 
\end{equation}
where $X_e(T)$ is the ionization fraction of free electrons in the cosmological plasma and we took for $\chi\ll 1$, $\Delta\mathcal I(\chi)=6/45$ to obtain expression \eqref{far-big-qed} and $\Delta\mathcal T_0(\chi)=-(1/4)\sqrt{3/2}\exp[-4/\chi]$ to obtain expression \eqref{far-big-pair}. By using numerics, we obtain for min $\{X_e^{1/4}(T)\,T^{-1/2}\}\simeq 4.34\times 10^{-3}$ (K$^{-1/2}$) at $T\simeq 1540.6$ K and min $\{X_e^{1/3}(T)\,T^{1/3}\}\simeq 0.24$ (K$^{1/3}$) at $T\simeq 57.22$ K. As we will see in what follows expression \eqref{far-big-pair} is satisfied for most values of the parameters.

Consider now generation of CMB circular polarization where its degree of polarization for $V_i=0$ is given by 
\begin{equation}\label{deg-circ-far}
P_C(T_0)=\frac{|-\mathcal S(T_0) Q_i-\mathcal W(T_0) U_i|}{\left[1+\mathcal Y(T_0)\right]I_i-\mathcal X(T_0) Q_i+\mathcal Z(T_0) U_i},
\end{equation}
where we have defined 
\begin{align}\label{xyz-def}
\mathcal S(T_0) &\equiv \int_{T_0}^{T_i}dT^{\prime}\,\tilde G(T^{\prime})\sin[\mathcal M_F(T^{\prime})], \quad \mathcal W(T_0)\equiv \int_{T_0}^{T_i}dT^{\prime}\,\tilde G(T^{\prime})\cos[\mathcal M_F(T^{\prime})], \nonumber \\ & \mathcal X(T_0)\equiv \int_{T_0}^{T_i}dT^{\prime}\,G_0(T^{\prime})\cos[\mathcal M_F(T^{\prime})], \quad \mathcal Z(T_0)\equiv \int_{T_0}^{T_i}dT^{\prime}\,G_0(T^{\prime})\sin[\mathcal M_F(T^{\prime})], \nonumber\\ & \mathcal Y(T_0)\equiv \int_{T_0}^{T_i}dT^\prime\,G_0(T^\prime)\cos[\mathcal M_F(T^\prime)]\, \int_{T^\prime}^{T_i}dT^{\prime\prime}\,G_0(T^{\prime\prime})\cos[\mathcal M_F(T^{\prime\prime})]+ \nonumber\\ & \int_{T_0}^{T_i}dT^\prime\,G_0(T^\prime) \sin[\mathcal M_F(T^\prime)]\,\int_{T^\prime}^{T_i}dT^{\prime\prime}\,G_0(T^{\prime\prime})\sin[\mathcal M_F(T^{\prime\prime})].
\end{align}
In most cases there are no known analytic expressions for $\mathcal X, \mathcal Y, \mathcal Z, \mathcal S$ and $\mathcal W$ unless one uses some approximations in order to simplify the form of integrands. In addition, in all functions defined in \eqref{xyz-def} appears $\mathcal M_F(T)$ which is proportional to the integral of the product $X_e(T)T^{1/2}$, where in general there is no known analytic expression for $X_e(T)$ which satisfies a complicated differential equation \cite{Weinberg:2008zzc}.

The expression \eqref{deg-circ-far} can be significantly simplified by considering the case when $G_0(T)=0$. This would happen when the decay of photons into milli-charged fermions occurs before decoupling time, so, only birefringence effect would occur at post decoupling epoch. Therefore, from \eqref{deg-circ-far} the degree of circular polarization for $G_0(T)=0$ becomes 
\begin{equation}\label{deg-circ-far-1}
P_C(T_0)=|-\mathcal S(T_0) (Q_i/I_i)-\mathcal W(T_0) (U_i/I_i)|.
\end{equation}
Now the expression for the degree of circular polarization has been quite simplified and we have to deal with only functions $\mathcal S$ and $\mathcal W$. Since in $\mathcal M_F(T)$ enters $X_e(T)$, there are essentially three possibilities to calculate $\mathcal S$ and $\mathcal W$. We may calculate them numerically given the numerical solution of $X_e$, or, we may substitute $X_e$ with its average value at the post decoupling epoch, or, we may look for values of the parameters in such a way to have $\mathcal M_F(T)<1$ and look for semi-analytic expressions.
 
 In this work we concentrate on the case when $\mathcal M_F(T)<1$, namely we look for values of the parameters that satisfy this condition. In this case we may approximate
 \begin{equation}\label{xyz-def-1}
 \mathcal S(T_0) \simeq \int_{T_0}^{T_i}dT^{\prime}\,\tilde G(T^{\prime})\,\mathcal M_F(T^{\prime}), \quad \mathcal W(T_0)\simeq \mathcal{\tilde G}(T_0),
\end{equation}
where $\mathcal M_F(T)$ is given by
\begin{equation}\label{M-def}
\mathcal M_F(T)=8.71\times 10^{25}\,\cos(\Phi)\,T_0^{-1/2}\left(\frac{\text{Hz}}{\nu_0}\right)^2\left(\frac{B_{e0}}{\text G}\right)\int_T^{T_i}dT^\prime\,X_e(T^\prime)\,T^{\prime 1/2}\quad (\text K^{-1}),
\end{equation}
see Ref. \cite{Ejlli:2016avx} for details.
In the expression for $\mathcal M_F(T_0)$ and $\mathcal S(T_0)$ appear the following integrals of the ionization fraction, which numerical values are given by
\begin{equation}\label{integrals}
\int_{T_0}^{T_i} dT^\prime X_e(T^\prime)\,T^{\prime 1/2}\simeq 1790.3\quad (\text K^{3/2}),\quad  \int_{T_0}^{T_i}dT^\prime\, T^{\prime 5/2}\int_{T^\prime}^{T_i} dT^{\prime\prime}\, X_e(T^{\prime\prime})\,T^{\prime\prime 1/2}\simeq 5.17\times 10^{14}\quad (\text K^5).
\end{equation}
In obtaining the numerical values of integrals in \eqref{integrals}, we used the numerical solution\footnote{The numerical solution of $X_e$ has been obtained by using cosmological parameters given by Planck collaboration \cite{Ade:2015xua}. } of the differential equation satisfied by $X_e(T)$ given in Ref. \cite{Weinberg:2008zzc}, in the temperature interval $57.22$ K $\leq T\leq T_i$ where $T_i=2970$ K is the temperature of the CMB at decoupling time. The temperature $T=57.22$ K corresponds to the start of the reionization epoch, where the ionization fraction would start adiabatically increase until complete ionization is reached at $T\simeq 21.8$ K or redshift $z\simeq 7$. The  evolution of the curve $X_e(T)$ in the temperature interval 21.8 K$\leq T\leq $ 57.22 K has been obtained by smooth interpolation of the solution of $X_e$ in the interval $57.22$ K $\leq T\leq T_i$, with $X_e\simeq 1$ in the interval 2.725 K$\leq T\leq 21.8$ K.

Now by using \eqref{integrals} in \eqref{xyz-def-1}, we get the following expression for the degree of circular polarization at the present time
\begin{equation}
P_C(T_0)=\left(\frac{\epsilon m_e}{m_\epsilon}\right)^4\sin^2(\Phi)\left|1.8\times 10^{26}\,\cos(\Phi)\left(\frac{\textrm{Hz}}{\nu_0}\right)\left(\frac{B_{e 0}}{\textrm{G}}\right)^3\,\frac{Q_i}{I_i}+2.7\times 10^{-3}\,\left(\frac{\nu_0}{\textrm{Hz}}\right)\left(\frac{B_{e 0}}{\textrm{G}}\right)^2\,\frac{U_i}{I_i}\right|.
\end{equation}
Now suppose that $\nu_0=33$ GHz (MIPOL observation frequency \cite{Mainini:2013mja}) and $B_{e0}=1$ nG, where for these values of the frequency and magnetic field amplitude obviously we have $\mathcal M(T_0)<1$. If we consider that at decoupling $U_i= r\,Q_i$, we get the following value for the degree of circular polarization
\begin{equation}\label{pol-far}
P_C(T_0)=\left(\frac{\epsilon m_e}{m_\epsilon}\right)^4\left(\frac{|Q_i|}{I_i}\right)\,\sin^2(\Phi)\left|5.45\times 10^{-12}\cos(\Phi)+8.91\times 10^{-11}\, r\right|.
\end{equation}
Using the MIPOL \cite{Mainini:2013mja} upper limit  on the degree of circular polarization $P_C(T_0)<7\times 10^{-5}$ and $|Q_i|/I_i\simeq 10^{-6}$, we get the following constraints
\begin{equation}\label{cons-1}
\begin{gathered}
1.44\times 10^{-7} \leq \epsilon< 8.27\times 10^{-6}\left|\frac{\cos(\Phi)}{\sin^2(\Phi)}\right|^{1/4}\, \text{for}\quad  7.4\times 10^{-2}\, \text{eV}\leq m_\epsilon\ll 5.1\times 10^5\,\epsilon\,\text{eV}\quad\text{or}\\ 8.27\times 10^{-6}\left|\frac{\cos(\Phi)}{\sin^2(\Phi)}\right|^{1/4} \leq \epsilon<1\quad \text{for}\quad 8930.12\,\epsilon\left|\frac{\cos(\Phi)}{\sin^2(\Phi)}\right|^{-1/4} \ll m_\epsilon \leq 5.1\times 10^5\,\epsilon\,\text{eV}.
\end{gathered}
\end{equation}
In obtaining the constraints in expression \eqref{cons-1}, we used the constraint $\epsilon<1$, constraints \eqref{chi-cond}-\eqref{critic-cond}, constraint \eqref{far-big-qed} and the lower limit on $m_\epsilon$ which comes from the fact that decay of photons into milli-charged fermions occurs before decoupling epoch.

\begin{figure*}[htbp!]
\centering
\mbox{
\subfloat[\label{fig:Fig7}]{\includegraphics[scale=0.65]{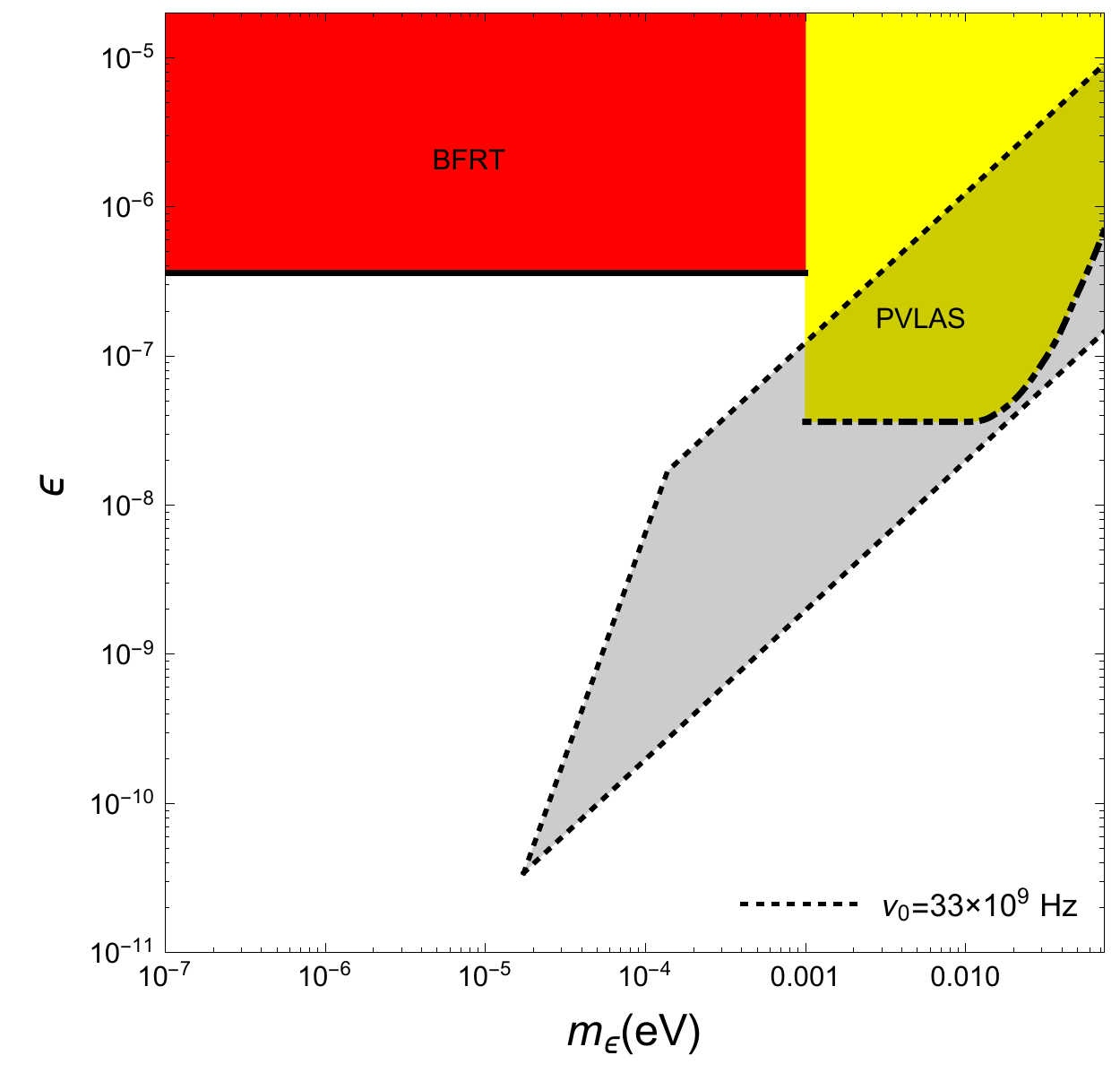}}\qquad
\subfloat[\label{fig:Fig8}]{\includegraphics[scale=0.65]{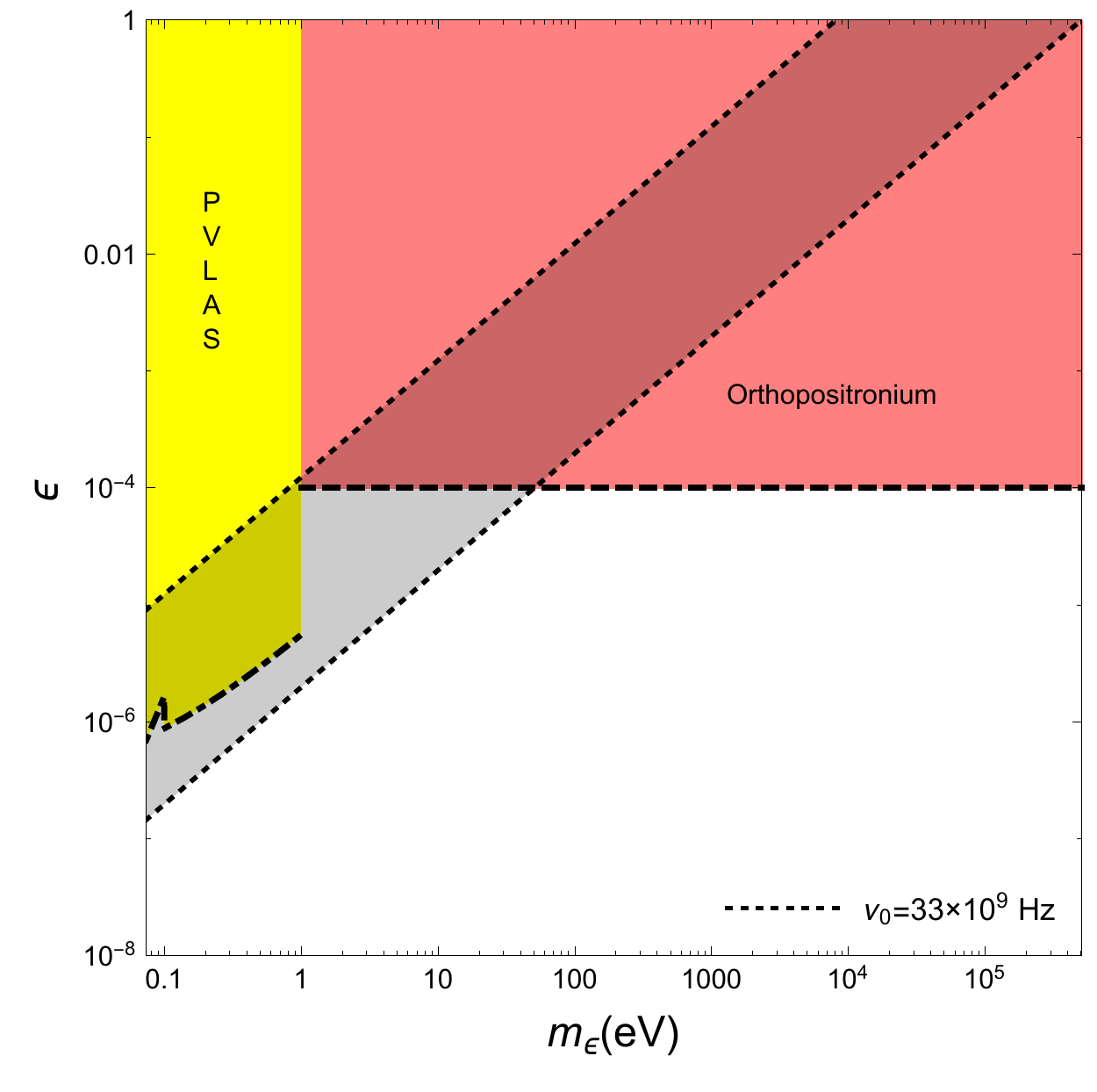}}}
\caption{In (a) \emph{allowed} region in grey within dotted lines, in the parameter space $\epsilon$ vs. $m_\epsilon$ for magnetic field amplitude $B_{e0}=$ 1 nG, $\Phi=\pi/4$, $\nu_0=33$ GHz and for photons that decay into milli-charged fermions at post decoupling for given frequency $\nu_0$, $m_\epsilon< 2.24\times 10^{-12}(\nu_0/\text{Hz})$ eV is shown. In (b) \emph{allowed} region in grey within dotted lines for same parameters $B_{e0}, \Phi$ and $\nu_0$ as in (a) for photons that do not decay into milli charged fermions at post decoupling for given frequency $\nu_0$, $m_\epsilon\geq 2.24\times 10^{-12}(\nu_0/\text{Hz})$ eV. In both plots, the \emph{exclusion} regions obtained by PVLAS, BFRT and invisible decays of Orthopositronium are also shown. }
\label{fig:Fig4a}
\end{figure*}

 In the case when $G_0(T_0)\neq 0$ for $T_0\leq T\leq T_i$, decay into milli-charged fermions would occur after decoupling time and would contribute to generation of CMB circular polarization. In this case the intensity of the CMB, $I(T)$, would change as a consequence of decay into milli-charged fermions. However, change in intensity is expected to be very small since in expressions $\mathcal Y(T_0), \mathcal X(T_0), \mathcal Z(T_0)$ do appear terms proportional to exponentials and Gamma functions of $\chi(T)\ll 1$, which for example when $\chi(T)\leq 0.01$ are extremely small value functions. Consequently, the functions $\mathcal Y(T_0), \mathcal X(T_0)$ and $\mathcal Z(T_0)$ have very small values in this regime for reasonable values of the pre-factors which enter into them. However, we must remind that this is an approximation which does not allow to explore the full range of parameters $\epsilon$ and $m_\epsilon$, especially those that satisfy $0.01\lesssim\chi(T)\ll 1$. So, for an order of magnitude estimate, we may approximate the term in the denominator of \eqref{deg-circ-far} with $I_i$ and the expression for $P_C(T_0)$ is still given by \eqref{deg-circ-far-1}. Now that $m_\epsilon$ is bounded from above, $m_\epsilon< 2.24\times 10^{-12}(\nu_0/\text{Hz})$ eV, we obtain the following limits for $m_\epsilon$ and $\epsilon$ from expression \eqref{pol-far} at $B_{e0}=1$ nG, for the MIPOL upper limit on $P_C(T_0)<7\times 10^{-5}$ at $\nu_0=33$ GHz
\begin{equation}\label{cons-3}
\begin{gathered}
3.42\times 10^{-11} \leq \epsilon <  1.47\times 10^{-8}\left|\frac{\cos(\Phi)}{\sin^2(\Phi)}\right|^{3/8} \quad \text{for}\quad   
R \leq \left(\frac{m_\epsilon}{\text{eV}}\right) \leq 5.1\times 10^5\, \epsilon\quad \text{or}\quad  \\
1.47\times 10^{-8}\left|\frac{\cos(\Phi)}{\sin^2(\Phi)}\right|^{3/8} \leq \epsilon< 1.44\times 10^{-7}\quad \text{for}\quad   8930.12\,\epsilon\left|\frac{\cos(\Phi)}{\sin^2(\Phi)}\right|^{-1/4} < \left(\frac{m_\epsilon}{\text{eV}}\right) \leq 5.1\times 10^{5}\,\epsilon\quad \text{or} \\
1.44\times 10^{-7} \leq \epsilon< 8.27\times 10^{-6}\left|\frac{\cos(\Phi)}{\sin^2(\Phi)}\right|^{1/4}\, \text{for}\quad  8930.12\,\epsilon\left|\frac{\cos(\Phi)}{\sin^2(\Phi)}\right|^{-1/4} < \left(\frac{m_\epsilon}{\text{eV}}\right)< 7.4\times 10^{-2}.
\end{gathered}
\end{equation}
In obtaining the constraints \eqref{cons-3} we used together with the condition $\epsilon<1$, the condition \eqref{con-0}, conditions \eqref{qed-dec-cond}-\eqref{critic-cond} and conditions \eqref{far-big-pair}-\eqref{far-big-qed}, all evaluated at $\nu_0=33$ GHz and $B_{e0}=1$ nG. In Fig. \ref{fig:Fig4a} the allowed regions (in grey) in the parameter space $\epsilon$ vs. $m_\epsilon$ for the solutions \eqref{cons-3}, Fig. \ref{fig:Fig7}, and for the solutions \eqref{cons-1}, Fig. \ref{fig:Fig8} are shown. In both plots we considered as a matter of example, photons propagating at an angle $\Phi=\pi/4$ against the direction of magnetic field. Excluding the value $\Phi=0$, where expressions \eqref{cons-1} and \eqref{cons-3} would be formally singular and the value $\Phi=\pi/2$, the allowed regions change very little for values of $0<\Phi<\pi/2$ when other parameters ($B_{e0}, \nu_0$) are considered fixed


 An interesting consequence of the case $\Phi\neq \pi/2$ and $\chi\ll 1$, there is rotation of the polarization plane of the CMB even when it is unpolarized at decoupling time. Indeed, we obtain the following expression for the rotation angle from solutions \eqref{pert-sol-par} for $V_i=0$
 \begin{equation}\label{rot-1}
\tan[2\psi(T_0)]\simeq -\frac{\left(\cos\left[\mathcal M_F(T_0)\right]\,\mathcal Z(T_0)-\sin\left[\mathcal M_F(T_0)\right]\mathcal X(T_0)\right) I_i+\sin\left[\mathcal M_F(T_0)\right]\,Q_i+\cos\left[\mathcal M_F(T_0)\right]\,U_i}{\left(\cos\left[\mathcal M_F(T_0)\right]\,\mathcal X(T_0)+\sin\left[\mathcal M_F(T_0)\right]\mathcal Z(T_0)\right)I_i-\cos\left[\mathcal M_F(T_0)\right]\,Q_i+\sin\left[\mathcal M_F(T_0)\right]\,U_i},
\end{equation}
where we neglected second order terms proportional to $Q_i$ and $U_i$ in expressions for $Q(T)$ and $U(T)$ in \eqref{pert-sol-par}. We may note from \eqref{rot-1} that the rotation angle depends on $G_0(T_0)$ and on the Faraday term $\mathcal M_F(T_0)$ since these terms are responsible for changing the intensity of the states $A_+$ and $A_\times$ and consequently there is rotation of the polarization plane. In the case when $G_0(T_0)=0$ in a given temperature interval, the rotation angle is only due to the Faraday effect in that interval. Moreover, even in the case when the CMB would be initially completely unpolarized $(Q_i=U_i=V_i=0)$ there is still rotation of the polarization plane. In order to have rotation of the polarization plane for initially unpolarized CMB, we must necessarily have $\mathcal X(T_0)\neq 0$, $\mathcal Z(T_0)\neq 0$ and $\mathcal M_F(T_0)\neq 0$. 

Typically, one looks for the rotation angle of polarization plane of the CMB at frequencies $\nu_0>10^{10}$ Hz and if in addition we consider that external magnetic field amplitude is $B_{e0}\leq 1$ nG, we may safely approximate $\mathcal M_F(T_0)< 1$. In such case we have
\begin{equation}\label{apro-1}
\cos\left[\mathcal M_F(T_0)\right]\simeq 1,\quad \sin\left[\mathcal M_F(T_0)\right]\simeq \mathcal M_F(T_0),\quad \mathcal Z(T_0)\simeq \int_{T_0}^{T_i}dT^{\prime}\,G_0(T^{\prime})\,\mathcal M_F(T^{\prime}), \quad \mathcal X(T_0)\simeq \mathcal G_0(T_0).
\end{equation}
Considering the fact that the rotation angle (in radians) of CMB polarization plane is in general a small quantity, we get from \eqref{rot-1} and \eqref{apro-1}
\begin{equation}\label{psi-1}
\delta\psi(T_0)\simeq \frac{1}{2(1+r^2)}\left[\frac{\left[\mathcal M_F(T_0)\mathcal G_0(T_0)-\mathcal Z(T_0)\right]\,I_i-\mathcal M_F(T_0)\,Q_i-r\,Q_i}{\left[\mathcal G_0(T_0)+\mathcal M_F(T_0)\mathcal Z(T_0)\right]I_i-Q_i+\mathcal M_F(T_0)\,r\,Q_i}-r\right],
\end{equation}
where we expressed $U_i=r Q_i$.

As mentioned in the paragraph above, the interesting thing about the case when $\Phi\neq \pi/2$ is that we have generation of linear polarization (at second order in perturabation theory) even in the case when the CMB would be initially unpolarized at the temperature $T_i$. Let us consider this situation further and concentrate for the moment on the rotation angle of the polarization plane which expression is given in \eqref{psi-1}. Since we already have calculated the expression for $\mathcal M_F(T_0)$ and $\mathcal G_0(T_0)$, the only left function in order to calculate $\delta\psi(T_0)$ in \eqref{psi-1} is $\mathcal Z(T)$. In general, $\mathcal Z(T)$ has no known analytic expression since in $\mathcal M_F(T)$ does appear $X_e(T)$. However, it is possible to find an analytic expression if one approximates $X_e$ as constant or more precisely by approximating it with its average value, $X_e(T)\simeq \bar X_e$ in a given temperature interval. By using expression \eqref{M-def}, we can write $\mathcal M_F(T)=\beta\, (T_i^{3/2}-T^{3/2})$ where $\beta\equiv 8.71\times 10^{25}\,(2/3)\cos(\Phi)\bar X_e\,T_0^{-1/2}\,(\text{Hz}/\nu_0)^2(B_{e0}/\text{G})$  (K$^{-1}$) and then obtain the following expression for $\mathcal Z(T_0)$
\begin{align}\label{Z-0}
\mathcal Z(T_0) &= -\frac{1}{8}\sqrt{\frac{3}{2}}\mathcal A\,\beta \left[3\,T_i^2\exp(-4/\chi_i)+ T_0^2\exp(-4/\chi_0)+2^{4/3}\mathcal B^{-2/3}\left(\Gamma\left(\frac{1}{3}, \frac{4}{\chi_i}\right)- \Gamma\left(\frac{1}{3}, \frac{4}{\chi_0}\right)\right)\right.\nonumber\\  & \left.
-4\,T_i^{3/2}\sqrt{T_0}\exp(-4/\chi_0)+2^{7/3}\mathcal B^{-1/6}T_i^{3/2}\left(\Gamma\left(\frac{5}{6}, \frac{4}{\chi_0}\right)- \Gamma\left(\frac{5}{6}, \frac{4}{\chi_i}\right)\right)\right],
\end{align}
where we expressed $\mathcal B$ as, $\mathcal B=\chi_i\,T_i^{-3}$, $\chi_i=\chi(T_i)$, $\chi_0=\chi(T_0)$ and considered for simplicity $T_\epsilon<T_0$. In the case when $T_\epsilon>T_0$ one must replace $T_0\rightarrow T_\epsilon$ in \eqref{Z-0} and other expressions which we derive below.


Now by using the expression for $\mathcal M_F(T_0)$, expressions \eqref{G-0-1} and \eqref{Z-0}, we get the following expression for the term in the numerator of \eqref{psi-1},
\begin{equation}\label{Z-2}
\begin{gathered}
\mathcal M_F(T_0)\mathcal G_0(T_0)-Z(T_0)= \frac{1}{8}\sqrt{\frac{3}{2}}\mathcal A\,\beta \,\left[T_i^2\exp(-4/\chi_i)-4T_i^{3/2}T_0^{1/2} \exp(-4/\chi_0)-4T_0^{3/2}T_i^{1/2} \exp(-4/\chi_i) + \right.\\ \left.
3T_0^{2} \exp(-4/\chi_0)+4T_0^{1/2}T_i^{3/2} \exp(-4/\chi_0)+2^{4/3}\chi_i^{-2/3}T_i^2\left(\Gamma\left(\frac{1}{3}, \frac{4}{\chi_0}\right)-\Gamma\left(\frac{1}{3}, \frac{4}{\chi_i}\right)\right) +\right.\\ \left.   2^{7/3}\chi_i^{-1/6}T_i^{1/2}T_0^{3/2}\left(\Gamma\left(\frac{5}{6}, \frac{4}{\chi_i}\right)-\Gamma\left(\frac{5}{6}, \frac{4}{\chi_0}\right)\right) \right],
\end{gathered}
\end{equation}
As we did for the function $\mathcal G_0(T_0)$, is the exponential term $\exp(-4/\chi_i)$ and Gamma functions that contain as second argument $4/\chi_i$ that give biggest contributions in expression \eqref{Z-2} since $\chi_0\ll \chi_i\ll 1$ for $T_0\ll T_i$. In addition we have that for $\chi_i\ll 1$, $\chi_i^{-1/6}\ll \chi_i^{-2/3}$. So, by doing these approximations, expression \eqref{Z-2} becomes 
\begin{equation}\label{Z-2-2}
\begin{gathered}
\mathcal M_F(T_0)\mathcal G_0(T_0)-Z(T_0)\simeq  \frac{1}{8}\sqrt{\frac{3}{2}}\mathcal A\,\beta \,T_i^2\,\exp(-4/\chi_i)\left[1- 4(T_0/T_i)^{3/2}-2^{4/3}\chi_i^{-2/3}\,\Gamma\left(\frac{1}{3}, \frac{4}{\chi_i}\right)\exp(4/\chi_i)\right.\\ \left. +2^{7/3}(T_0/T_i)^{3/2}\chi_i^{-1/6}\,\Gamma\left(\frac{5}{6}, \frac{4}{\chi_i}\right)\exp(4/\chi_i)\right].
\end{gathered}
\end{equation}
By doing same reasoning as done above in deriving \eqref{Z-2-2}, the expression in the denominator of \eqref{psi-1} becomes
 \begin{equation}\label{Z-3}
 \begin{gathered}
\mathcal G_0(T_0)+\mathcal M_F(T_0)\mathcal Z(T) \simeq -\frac{1}{2}\sqrt{\frac{3}{2}}\mathcal A\,T_i^{1/2}\,\exp(-4/\chi_i)\left[1-2^{1/3}\mathcal \chi_i^{-1/6}\,\Gamma\left(\frac{5}{6}, \frac{4}{\chi_i}\right)\exp(4/\chi_i)\right]-\frac{1}{8}\sqrt{\frac{3}{2}}\mathcal A\,\beta^2\,T_i^{7/2}\\ \times \exp[-4/\chi_i]\,   \left[3-3(T_0/T_i)^{3/2} -2^{7/3}\chi_i^{-1/6}\Gamma\left(\frac{5}{6}, \frac{4}{\chi_i}\right)\exp[4/\chi_i]+ 2^{4/3}\chi_i^{-2/3}\Gamma\left(\frac{1}{3}, \frac{4}{\chi_i}\right)\exp[4/\chi_i]\right.\\
\left. + 2^{7/3}\chi_i^{-1/6}(T_0/T_i)^{3/2}\Gamma\left(\frac{5}{6}, \frac{4}{\chi_i}\right)\exp[4/\chi_i] - 2^{4/3}\chi_i^{-2/3}(T_0/T_i)^{3/2}\Gamma\left(\frac{1}{3}, \frac{4}{\chi_i}\right)\exp[4/\chi_i]  \right].
\end{gathered}
\end{equation}
 In the case when $T_0\ll T_i$, the terms proportional to $(T_0/T_i)^{3/2}$ in expression \eqref{Z-3} can be neglected with respect to terms that do not contain $(T_0/T_i)^{3/2}$. So, after neglecting these terms, let us define $\eta$ and $\xi$ as
 \begin{equation}\nonumber
 \begin{gathered}
\eta\equiv \frac{3+2^{4/3}\mathcal \chi_i^{-2/3}\,\Gamma\left(\frac{1}{3}, \frac{4}{\chi_i}\right)\,\exp[4/\chi_i]-2^{7/3}\mathcal \chi_i^{-1/6}\,\Gamma\left(\frac{5}{6}, \frac{4}{\chi_i}\right)\,\exp[4/\chi_i]}{1-2^{1/3}\mathcal \chi_i^{-1/6}\,\Gamma\left(\frac{5}{6}, \frac{4}{\chi_i}\right)\,\exp[4/\chi_i]},\\
\xi\equiv \frac{1-2^{4/3}\mathcal \chi_i^{-2/3}\,\Gamma\left(\frac{1}{3}, \frac{4}{\chi_i}\right)\,\exp[4/\chi_i]}{1-2^{1/3}\mathcal \chi_i^{-1/6}\,\Gamma\left(\frac{5}{6}, \frac{4}{\chi_i}\right)\,\exp[4/\chi_i]},
\end{gathered}
\end{equation}
then by using expressions \eqref{Z-2-2}-\eqref{Z-3} into \eqref{psi-1}, we get the following expression for the rotation angle for unpolarized CMB at $T_i$ ($Q_i=U_i=V_i=0$ and $r=0$)
\begin{equation}\label{unpol-far}
\delta\psi(T_0)\simeq -\frac{\beta T_i^{3/2}}{8}\left[\frac{\xi}{1+(\beta T_i^{3/2}/2)^2\,\eta}\right].
\end{equation}

Expression \eqref{unpol-far}, which has been derived in the case when $\mathcal M_F(T_0)<1$, is of great importance since it implies that there is rotation of the CMB polarization plane, as anticipated earlier, even in the case when it is initially unpolarized and in addition the rotation angle is proportional to the Faraday term, $\mathcal M_F(T_0)=\beta (T_i^{3/2}-T_0^{3/2})\simeq \beta T_i^{3/2}$ for $T_0\ll T_i$. Moreover, in expression \eqref{unpol-far} there are correction terms proportional to $\eta$ and $\xi$. Their values essentially depend on $\chi_i$ and by using the property $\Gamma(s, x)\,e^{x}/x^{s-1}\rightarrow 1$ for $x\rightarrow \infty$, we have essentially that $\eta$ and $\xi$ are undetermined for $\chi_i\rightarrow 0$, namely $\eta, \xi \rightarrow 0/0$.  In case when $\chi_i$ is extremely small ($\chi_i\rightarrow 0$) one must include other mixed terms proportional to $(T_0/T_i)^{3/2}$ that we neglected above, in expressions of $\eta$ and $\xi$. However, we are not interested in such extremely small values of $\chi_i$ since we want to explore a vast range in the parameter space $\epsilon, m_\epsilon$. Now, all told let us consider for example the case when $\chi_i\simeq 10^{-2}$ as we did in previous sections and from definition of $\eta$ and $\xi$, we get respectively $\eta \simeq 4.96\times 10^{-3},\,\xi\simeq 4$ . 

Consider now the case that magnetic fields exist prior decoupling time, say when the temperature is about $T_i\simeq 10^4$ K where according to standard cosmology the CMB is unpolarized. Also suppose that we observe the CMB today at frequency $\nu_0=53$ GHz. For these values of the parameters we get $\mathcal M_F(T_0)\simeq 8.12\times 10^{9}\cos(\Phi)\,(B_{e0}/\text {G})$, where we used the average value of $\bar X_e\simeq 0.65$ in the temperature interval $T_0\leq T\leq T_i$ for $T_i=10^4$ K. If we consider for example that $B_{e0}\leq 0.1$ nG, we get $\mathcal M_F(T_0)\leq 0.8\cos(\Phi)<1$. Now since $\eta\simeq 4.96\times 10^{-3}$ and $\xi\simeq 4$ for $\chi_i\simeq 0.01$, we have from expression \eqref{unpol-far} that essentially $\delta\psi(T_0)\simeq \mathcal -M_F(T_0)/2$. The latter condition tells us that as far as $2\delta\psi(T_0)<1$, the condition $\mathcal M_F(T_0)<1$ is automatically satisfied. We will see below that for current limits on $\delta\psi(T_0)$ set by experiments, this is indeed the case. 

\begin{figure*}[htbp!]
\centering
\mbox{
\subfloat[\label{fig:Fig9}]{\includegraphics[scale=0.65]{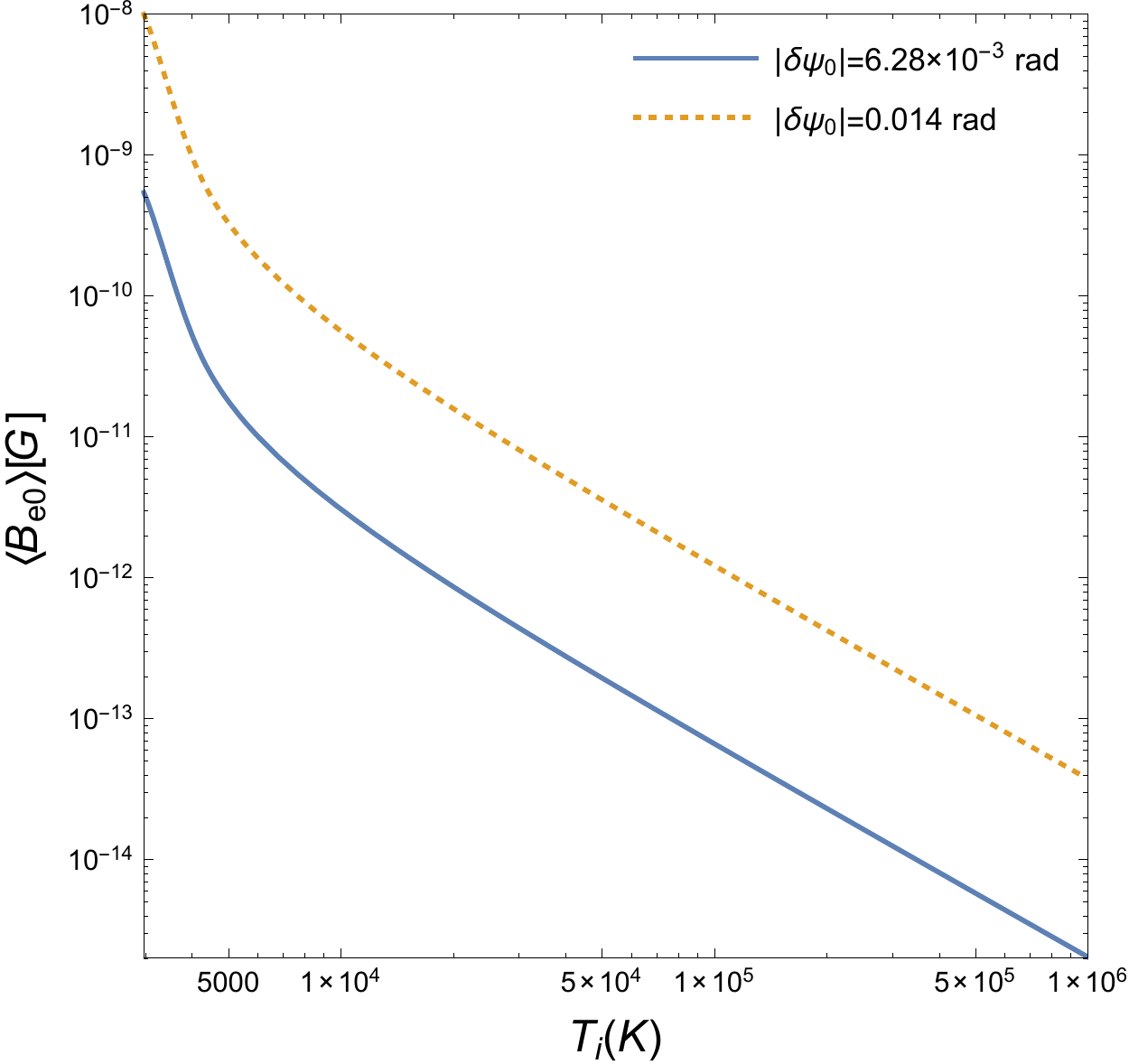}}\qquad
\subfloat[\label{fig:Fig10}]{\includegraphics[scale=0.65]{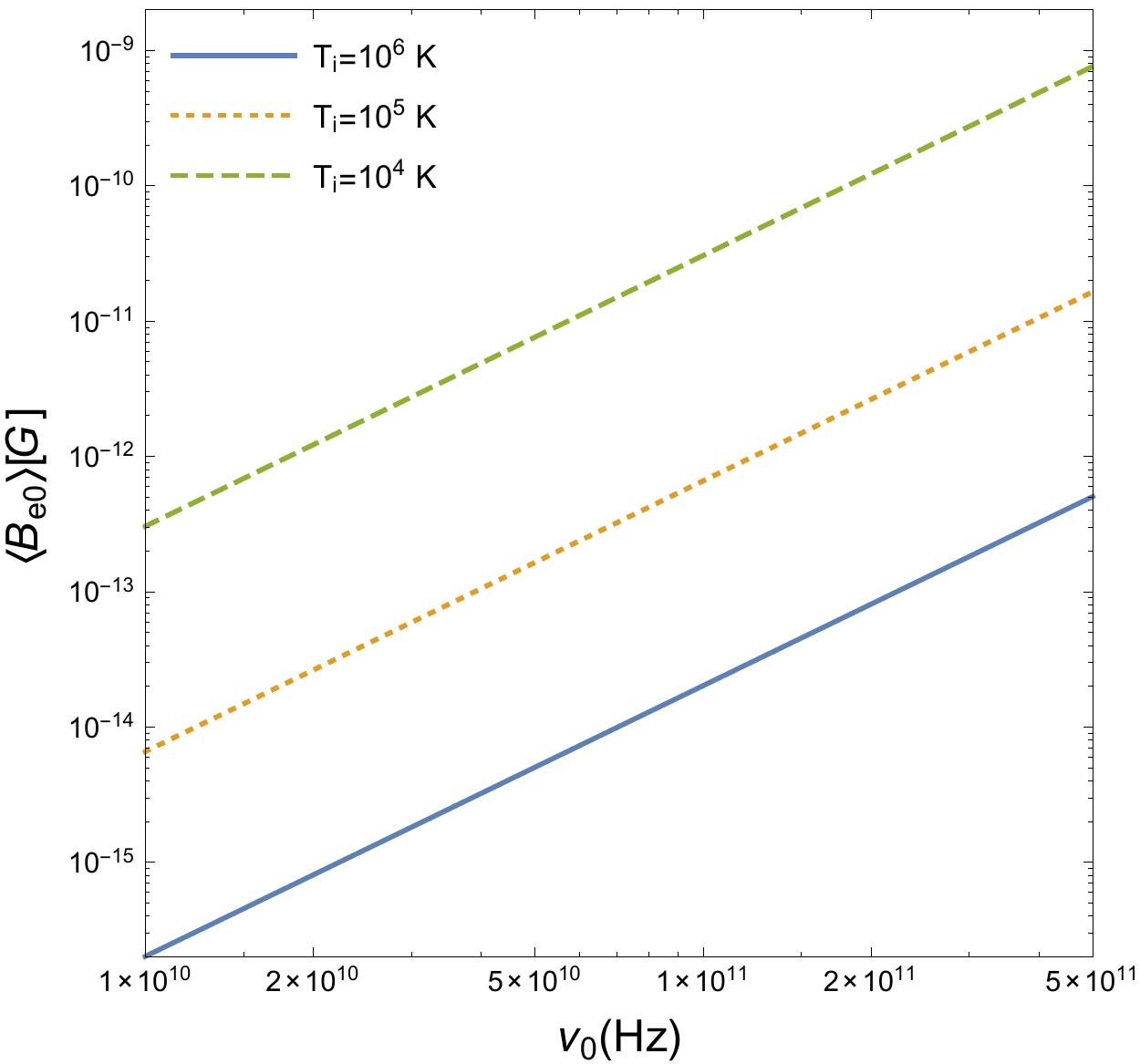}}}
\caption{In (a) plots of the average values over $\Phi$ of the magnetic field amplitudes at present, $\langle B_{e0}\rangle$ as a function of the initial temperature $2970\, \text{K}\leq T_i\leq 10^6\, \text{K}$ of generation of CMB polarization for the limits on $|\delta\psi(T_0)|=6.28\times 10^{-3}$ (rad) found by WMPA9 \cite{Hinshaw:2012aka} at $\nu_0=53$ GHz (solid line) and $|\delta\psi(T_0)|=0.014$ (rad) found by QUAD \cite{Wu:2008qb} at $\nu_0=150$ GHz (dotted line) are shown. In (b) plots of the average values over $\Phi$ of the magnetic field amplitudes at present, $\langle B_{e0}\rangle$ as a function of observation frequency $\nu_0$ for $|\delta\psi(T_0)|=1\, (\text{deg})$ or equivalently 0.0174 (rad) and initial CMB polarization generation temperatures $T_i=10^4$ K (dashed line), $T_i=10^5$ K (dotted line) and $T_i=10^6$ K (solid line) are shown. }
\label{fig:Fig5a}
\end{figure*}

If we use the current limit on $\delta\psi(T_0)=-6.28\times 10^{-3}$ (rad) obtained by WMPA9 at $\nu_0=53$ GHz, we get the following limit on $B_{e0}\simeq 1.54\times 10^{-12}\cos^{-1}(\Phi)$ G. This limit on magnetic field amplitude has been derived on the assumption that once generation of polarization starts at $T_i$ due to milli-charged fermion vacuum polarization, the contribution of other processes that create photons and which might destroy polarization, is negliglible. In Fig. \ref{fig:Fig5a} average values over observation angle $\Phi$ of present day magnetic field amplitudes for unpolarized CMB at initial temperature $T_i$ are shown. Our plots have been obtained by using expression \eqref{unpol-far} in the case when $\mathcal M_F(T_0)<1$, $\chi_i=10^{-2}$ and $\xi\simeq 4$. We can observe from \ref{fig:Fig9} that higher is the initial temperature when generation of CMB polarization starts, lower is the average value of $B_{e0}$ for given current limits on $\delta\psi(T_0)$ set by experiments.

 \subsection{Generation of polarization in case $\chi\gg 1$.}

 In the case when $\chi\gg 1$, the expression for the degree of circular polarization obviously is still given by \eqref{deg-circ-far}. However, now the expressions for the functions $\mathcal S, \mathcal W, \mathcal Y, \mathcal X$ and $\mathcal Z$ change with respect to the case $\chi\ll 1$. Even in the case when $\Phi\neq \pi/2$, the contribution of milli-charged fermion vacuum polarization to the circular polarization is very small as we explicitly have checked. We saw similar situation in Sec. \ref{sec:subsection-1}, in the case when $\Phi=\pi/2$ and $\chi\ll 1$. However, even though it is of not practical interest to study circular polarization in the case when $\Phi\neq\pi/2$ and $\chi\gg 1$, it is interesting to study the effect of milli-charged fermion vacuum polarization to the rotation angle of CMB polarization plane. 
 
Consider again the case when $\mathcal M_F(T_0)<1$, where the expression for the rotation angle of polarization plane is still given by expression \eqref{psi-1}. Now we need to calculate the expressions for $\mathcal X(T_0)\simeq \mathcal G_0(T_0)$ and $\mathcal Z(T)$. The expression for $\mathcal X(T_0)$ in case when $\chi\gg 1$ is given by
\begin{equation}\label{X}
\mathcal X(T_0)\simeq \mathcal G_0(T_0)  = \mathcal A \int_{T_0}^{T_i} T^{\prime -1/2}\Delta \mathcal T_0[\chi(T^\prime)]\, dT^{\prime}=  \frac{2^{4/3}\sqrt{3}\,\Gamma^2(2/3)}{7\sqrt{\pi}\,\Gamma(7/6)}\,\mathcal A\mathcal B^{-1/3}\,(T_i^{-1/2}-T_0^{-1/2}), 
\end{equation}
where we made use of expression \eqref{DI-0} in case of $\chi\gg 1$. The expression for $\mathcal Z(T_0)$ is given by
  \begin{equation}\label{Z-4}
\mathcal Z(T_0)\simeq  \mathcal A\beta \int_{T_0}^{T_i} T^{\prime -1/2}\Delta \mathcal T_0[\chi(T^\prime)](T_i^{3/2}-T^{\prime 3/2})\, dT^{\prime}= -\frac{2^{1/3}\sqrt{3}\,\Gamma^2(2/3)}{7\sqrt{\pi}\,\Gamma(7/6)}\, \mathcal A\mathcal B^{-1/3}\,\beta (-3T_i+T_0+2T_i\sqrt{T_i/T_0}),
 \end{equation}
 where again made use of expression \eqref{DI-0} in case of $\chi\gg 1$. In both expressions \eqref{X}-\eqref{Z-4} we considered for simplicity the case when $T_\epsilon<T_0$. In case when $T_\epsilon>T_0$, one must replace the lower limit of integration in expressions \eqref{X}-\eqref{Z-4} with $T_0\rightarrow T_\epsilon$.

 Let us consider again as we did in the previous section that the CMB is unpolarized at time $T_i$ which does not necessary coincide with decoupling temperature. By using expressions \eqref{X}-\eqref{Z-4} in \eqref{psi-1}, we get the following expression for the rotation angle of the polarization plane 
\begin{equation}\label{rot-2}
\delta\psi(T_0)= -\frac{1}{2}\left[\frac{\beta\left(T_i-3 T_0+2T_0\sqrt{T_0/T_i}\right)}{2\,\left(T_i^{-1/2}-T_0^{-1/2}\right)+\beta^2 \left(3T_i^{5/2}+T_0T_i^{3/2}-2T_i^{3}T_0^{-1/2}-3T_iT_0^{3/2}+T_0^{5/2}\right)}\right].
\end{equation}
In the case when $T_0<T_\epsilon<T_i$, one must replace $T_0$ in \eqref{rot-2} with $T_\epsilon$. It is interesting to note that in expression \eqref{rot-2} the rotation angle does not explicitly depend neither on $\epsilon$ nor on $m_\epsilon$. This is because we considered the case when $T_\epsilon<T_0$. In the opposite case there is a dependence on these parameters. 
However, since we are in the regime where $\chi(T)\gg 1$, the condition given in expression \eqref{con-6} must be satisfied and as we already have discussed this occurs for very low mass milli-charged fermions. It is quite straightforward to check that for such low mass milli-charged fermions, we have always $T_\epsilon\ll T_0$ for reasonable values of $\nu_0$ and $B_{e0}$. 
Consequently, in what follows we consider only the case when $T_\epsilon< T_0$.

Since here we are interested in the case when $T_i\gg T_0$, we can safely approximate expression \eqref{rot-2} in this regime by
\begin{equation}\label{rot-3}
\delta\psi(T_0)\simeq  \frac{1}{4}\left[\frac{\beta\, T_i\,T_0^{1/2}}{1+\beta^2 \,T_i^{3}}\right].
\end{equation}
The second term in the denominator of expression \eqref{rot-3} is exactly $\mathcal M_F^2(T_0)$ for $T_0\ll T_i$ and since we are in the regime where $\mathcal M_F(T_0)<1$, we have essentially to good accuracy $\beta^2 T_i^3\ll 1$. Expression \eqref{rot-3} can also be written in the following form $\beta T_i^{3/2}\simeq 4|\delta\psi(T_0)|(T_i/T_0)^{1/2}<1$ which in turn can be written as a condition on the initial temperature $T_i<T_0/(16\delta\psi^2(T_0))$. So, as far as the initial temperature satisfies the latter condition, we have automatically that $\mathcal M_F(T_0)<1$. Consider now the case when $T_i$ corresponds to a temperature when the CMB is in thermal equilibrium (before decoupling time) and also suppose that magnetic field exist at this temperature where would start generation of CMB polarization with rotation of the polarization plane. In this case, the expression for the average value over $0\leq \Phi\leq 2\pi$ of the magnitude of present day magnetic field which would rotate the polarization plane by $\delta\psi(T_0)$ until present time, at observation frequency $\nu_0$ and initial starting temperature $T_i$, is given by
\begin{equation}
\left(\frac{\langle B_{e0}\rangle }{\text{G}}\right)\simeq 1.37\times 10^{-25} \bar X_e^{-1}|\delta\psi(T_0)|\left(\frac{\nu_0}{\text{Hz}}\right)^2\,T_i^{-1}\quad (\text{K}).
\end{equation}
Consider for example that $|\delta\psi(T_0)|=6.28\times 10^{-3}$ (rad) at $\nu_0=53$ GHz, where we get from $T_i<T_0/(16\, \delta\psi^2(T_0))$, $T_i<4318.4$ K. If we take $T_i\simeq 4000$ K as starting temperature of generation of CMB polarization, we get the following value for $\langle B_{e0}\rangle=1.21\times 10^{-9}$ G. Here we used $\bar X_e\simeq 0.5$ for the average value in the interval $T_0\leq T\leq T_i$. Again as in the previous section, this calculation does not take into account any mechanism that might destroy polarization and which pushes the system toward thermal equilibrium.

 \section{Conclusions}
 \label{sec:6}

 In this work we have studied consequences of milli-charged fermion vacuum polarization in cosmic magnetic fields on the CMB polarization. This effect generates elliptic polarization of the CMB depending on the circumstances and even in the case when the CMB would be initially unpolarized. The effect studied in this work belongs to the category of magneto-optic effects, where there is close similarity with vacuum polarization due to electron/positron pair. However, as we have seen in this work the magnitude of birefringence and dichroism effects  due to milli-charged fermion vacuum polarization, can be much larger than birefringence and dichroism effects caused by standard vacuum polarization.
 
In order to compare our results with experimentally CMB measured or constrained quantities, we worked with the Stokes parameters and solved their equation of motion in expanding universe. Typically for this kind of problem there are not known analytic solutions of the equations of motion, unless in some particular cases, and one must use perturbation theory. The use of perturbation theory is strictly related to the magnitude of $\epsilon, m_\epsilon$ and $\Phi$, if one fixes $B_{e0}$ and $\nu_0$. Especially, the angle of observation $\Phi$ plays an important role because depending on its value with respect to the external magnetic field, we can solve the equations of motion exactly for $\Phi=\pi/2$ or use perturbation theory for $\Phi\neq \pi/2$. In addition, all quantities of interest such as the degree of circular polarization $P_C(T_0)$ and/or the rotation angle of the polarization plane $\delta\psi(T_0)$ depend explicitly on $\Phi$. This fact essentially means there in not uniformity across the sky of these quantities.

 In the case when $\Phi=\pi/2$, we found exact solutions of the Stokes parameters and estimated the degree of circular polarization $P_C(T_0)$ and $\delta\psi(T_0)$. The magnitude of these quantities depends on $B_{e0}, \nu_0, \Phi, m_\epsilon$ and $\epsilon$. In this work, we usually  fixed the magnitude of external magnetic field at $B_{e0}=1$ nG and let $\nu_0$ assume several values which mostly correspond with working frequencies of several experiments. Therefore, the only independent parameters remain $m_\epsilon$ and $\epsilon$. Another important factor is the mass of milli-charged fermion, $m_\epsilon$, since based on its value, we have either photons decay into milli-charged fermions  before or after decoupling poch. In the case when $m_\epsilon\geq 2.24\times 10^{-12}(\nu_0/\text{Hz})$ eV, photons decay before decoupling while in the opposite case, they decay after decoupling epoch for given observation frequency $\nu_0$.

 The expression of the degree of circular polarization $P_C(T_0)$ given in \eqref {deg-circ-pol-0}, in the case of $\Phi=\pi/2$, contains trigonometric functions and in principle can have multiple solutions in terms of $m_\epsilon$ and $\epsilon$ for given value or upper limit on $P_C(T_0)$. Usually, the value of $P_C(T_0)$ depends on the ratio $\epsilon/m_\epsilon$ for fixed values of $\nu_0, B_{e0}$ and $\Phi$. In the case when milli-charged fermions decay either before or after decoupling epoch for given observation frequency $\nu_0$, we found values of parameter space $\epsilon$ and $m_\epsilon$ allowed by our constraints and not excluded by experiments as shown in Figs. \ref{fig:Fig1a} and \ref{fig:Fig2a}. The most important factor which affects the circular polarization, for fixed $B_{e0}$ and $\nu_0$, is the ratio $\epsilon/m_\epsilon$ which does appear in expression of $P_C(T_0)$. 
 
 As shown in expression \eqref{unity-cond} and in Fig. \ref{fig:Fig3a}, the degree of circular polarization at present time could be close or even equal to present value of degree of linear polarization. If one fixes the magnetic field amplitude at order of $\sim$ nG and the value of $|r|$, the range of interesting values of $\epsilon/m_\epsilon$ varies with observation frequency. Higher is the value of $\epsilon/m_\epsilon$, higher is the value of expected degree of circular polarization at observation frequency $\nu_0$. Higher is the value of observation frequency for fixed $\epsilon/m_\epsilon$, higher is the value of expected degree of circular polarization. If one assumes that degree of circular polarization is smaller than degree of linear polarization, an optimistic \emph{future detection} range value for CMB circular polarization would be $P_C(T_0)\simeq 10^{-10}-10^{-6}$, given that present upper value set by MIPOL experiment is $P_C(T_0)<7\times 10^{-5}$. For example, if one observes the CMB say at frequency $\nu_0\sim 10^{10}$ Hz, in order to have $P_C(T_0)\simeq 10^{-10}-10^{-7}$ for $B=1$ nG and $|r|=0.1$, we must have $\epsilon/m_\epsilon\simeq (1.5 \times 10^{-4}-8\times 10^{-4})$ eV$^{-1}$.
 
 Obviously, there are values of $\epsilon$ and $m_\epsilon$ in the allowed region in Fig. \ref{fig:Fig4} for $\nu_0\sim 10^{10}$, that give a ratio $\epsilon/m_\epsilon\simeq (1.5 \times 10^{-4}-8\times 10^{-4})$ eV$^{-1}$. In principle, one may use for example the model depended limit on $\epsilon$ from BBN in order to find limits on $m_\epsilon$. The relation between effective neutrino species and $\epsilon$ is given by $\Delta N_\text{eff}=0.69\times 10^{17}\,\epsilon^2$ for no elastic scattering, see Ref. \cite{Davidson:2000hf}. If one uses the current limit on $\Delta N_\text{eff}$ obtained by Planck collaboration \cite{Ade:2015xua}, $\Delta N_\text{eff}=3.15$, we get the following BBN limit on $\epsilon\lesssim 6.75\times 10^{-9}$. If for example $\epsilon/m_\epsilon\simeq 1.5\times 10^{-4}$ eV$^{-1}$, by using the BBN upper limit on $\epsilon$ we get $m_\epsilon\lesssim 4.5\times 10^{-5}$ eV, while if $\epsilon/m_\epsilon\simeq 8\times 10^{-4}$ eV$^{-1}$,  we get $m_\epsilon\lesssim 8.43\times 10^{-6}$ eV. Now, one can easily verify that $\epsilon\lesssim 6.75\times 10^{-9}$ and $m_\epsilon\lesssim 4.5\times 10^{-5}$ eV is within the allowed region given in Fig. \ref{fig:Fig4} and not excluded by experiments, while the point $\epsilon\lesssim 6.75\times 10^{-9}$ and $m_\epsilon\lesssim 8.43\times 10^{-6}$ eV is outside the allowed region in Fig. \ref{fig:Fig4} for $\nu_0\sim 10^{10}$ Hz but allowed by experiments. This simple estimate suggest that BBN limit on $\epsilon$, implies, for our parameter space allowed by our constraints, an upper limit on $m_\epsilon\lesssim 4.5\times 10^{-5}$ eV and an upper limit on $P_C(T_0)\lesssim 10^{-10}$ for $\nu_0\sim 10^{10}$ Hz. One must interpret these results with caution since BBN limit on $\Delta N_\text{eff}$ is subjected to several uncertainties due to primordial abundance of light elements.
    
 In the case when $\Phi\neq \pi/2$, the Faraday effect gives significant contribution to generation of circular polarization. However, the expression for the degree of circular polarization becomes analytically more difficult to treat since in the expression for $\mathcal M_F$ does appear $X_e(T)$. The situation gets simplified in the case when $\mathcal M_F<1$, which essentially happens for $\nu_0\geq 10^{10}$ Hz and $B_{e0}\leq 10^{-9}$ G at post decoupling epoch. The Faraday effect does not generate circular polarization by itself, but it contributes to circular polarization due to its coupling with terms that generate birefringence effects as shown in the expression for $V(T)$ in \eqref{pert-sol-par}. In the case when $\mathcal M_F<1$, the contribution of Faraday effect to circular polarization is proportional to $Q_i$ as shown in expression \eqref{pol-far}. By using the MIPOL upper limit on $P_C(T_0)$, in Fig. \ref{fig:Fig4a}, the allowed regions in grey are shown. Even in the case when $\Phi\neq \pi/2$, apply the same discussions done above for the case $\Phi= \pi/2$, namely higher is the value of the ratio $\epsilon/m_\epsilon$, higher is the value of expected degree of circular polarization.

In addition to generation of CMB circular polarization, we also studied consequences of milli-charged fermion vacuum polarization on the rotation of the polarization plane of CMB. We were able to relate the ratio $\epsilon/m_\epsilon$ with rotation angle of the polarization plane, $\delta\psi(T_0)$, as shown in expression \eqref{rot-angle-1} in the case when $\Phi=\pi/2$. The interesting fact about expression \eqref{rot-angle-1}, there is a linear relation between $\epsilon$ and $m_\epsilon$ once other parameters are fixed.  If one uses the current value on $\delta\psi(T_0)\simeq -6.28\times 10^{-3}$ (rad) obtained by WMAP9 at $\nu_0=53$ GHz and $r=0.1$ and $B_{e0}=1$ nG, we would get $\epsilon/m_\epsilon\simeq 4.76\times 10^{-4}$ eV$^{-1}$. For this value of $\epsilon/m_\epsilon$, the value of acquired degree of circular polarization today would be $P_C(T_0)\simeq 5\times 10^{-8}$, namely two orders of magnitude smaller than present value of degree of linear polarization. Another important fact which is worth to mention is that as far as $|\delta\psi(T_0)|<|r|/4(1+r^2)$, the degree of circular polarization does not explicitly depend on observation frequency $\nu_0$ and $B_{e0}$ as shown in \eqref{rel-PC-psi}. So, even in the case when $r=0.01$, for same values of $B_{e0}$ and $\delta\psi(T_0)$, we would get $P_C(T_0)\simeq 1.5\times 10^{-8}$. This fact would suggest that for a canonical value of external magnetic field of order $\sim$ nG, the current limit on $\delta\psi(T_0)$ found by WMPA9 would be consistent with a limit on degree of circular polarization of the order $P_C(T_0)\sim 10^{-8}$. However, even this result must be interpreted with caution since the current limit on $\delta\psi(T_0)$ is subjected to statistical and systematic uncertainties and it has been found under the hypothesis of uniform rotation across the sky.
 
In the case when $\Phi\neq \pi/2$, the expression for the rotation angle is given by expression \eqref{rot-1}. In this regime we studied essentially the effect of milli-charged fermion vacuum polarization in the cases when $\chi\ll 1$, $\chi\gg 1$ and when $\mathcal M_F(T_0)<1$. In the case when $\chi\ll 1$ it turns out that rotation angle of the polarization plane is proportional to the term corresponding to the Faraday effect while in the case when $\chi\gg 1$ the situation is different. For $\chi\ll 1$, we derived expression \eqref{unpol-far} in the case when the CMB would be initially unpolarized at temperature $T_i$. Here the contribution of dichroism effect caused by milli-charged fermion vacuum polarization is encoded in the parameters $\eta$ and $\chi$. The interesting fact is that the CMB would have rotated its polarization plane starting before decoupling epoch if magnetic field was present at that time. Higher is the temperature when generation of CMB polarization would start, lower would be the magnetic field amplitude in order to generate rotation of polarization plane with current limit on $\delta\psi(T_0)$ given by experiments. In the case case when $\chi\gg 1$, the rotation angle of the polarization plane does not expend explicitly on milli-charged fermions parameters in the case when $T_\epsilon<T_0$ as shown in expression \eqref{rot-2}. 

 
Last, there are four important considerations which deserve attention. The first one is that for $\chi\ll 1$ and $\Phi=\pi/2$, we considered the cases when $\mathcal G_0(T_0)$ is either zero or much less than one, which essentially mean that dichroism effect due to decay of photons into milli-charged fermions is small and its contribution to $P_C(T_0)$ and/or $\delta\psi(T_0)$ is marginal. So, our results presented regarding generation of circular polarization and rotation of the polarization plane of the CMB, for $\chi\ll 1$ and $\Phi=\pi/2$, are essentially consequence of birefringence effect of the CMB in cosmic magnetic fields. The case when values of $\mathcal G_0(T_0)$ are of order of unity or higher for $\chi\ll 1$ and $\Phi=\pi/2$ have not been studied. 

The second consideration is related to the nature of cosmic magnetic field(s). In this work, as anticipated in Sec. \ref{sec:1}, the magnetic field has been assumed to be slowly varying function in space and time \emph{with respect} to the Compton wavelength of the milli-charged fermion and to the corresponding time interval. This is because the expressions for the photon polarization tensor and derived quantities like indexes of refraction are calculated in the regime $|\partial_\mu F_{\sigma\rho}| \ll m_e \,|F_{\sigma\rho}|$, see Refs. \cite{Heisenberg:1935qt}-\cite{Tsai:1974fa} for details, which translated to the case of milli-charged fermions is $|\partial_\mu F_{\sigma\rho}| \ll m_\epsilon \,|F_{\sigma\rho}|$. As far as the latter condition is valid, the expressions for photon polarization tensor also can be applied to the case of stochastic magnetic fields, which case has not been studied in this work. In an expanding universe and in the presence of only external magnetic field, the condition $|\partial B_e^i(\bs x, t)/\partial t|\ll m_\epsilon |B_e^i(\bs x, t)|$ translates to $H^{-1}(t)\gg 2/m_\epsilon=3.95\times 10^{-5}(\text{eV}/m_\epsilon)$ cm, while the condition $|\partial B_e^i(\bs x, t)/\partial \bs x|\ll m_\epsilon |B_e^i(\bs x, t)|$ translates to $l_B\gg 2/m_\epsilon=3.95\times 10^{-5}(\text{eV}/m_\epsilon)$ cm, where $B_e^i$ is the $i$-th component of $\bs B_e$ and $l_B(t)$ is the variation scale in space of external magnetic field. These conditions are satisfied for wide range of values of $m_\epsilon$ for values of $H(t)$ at post decoupling epoch. 

Third consideration is that expressions found for the Stokes parameters can be used to construct the multipole correlation functions $C_l^{VV}$ and $C_l^{VT}$ in the case of circular polarization, which might be more useful quantities than $P_C(T_0)$ in some circumstances. Since the $V$ parameter depends on $\epsilon, m_\epsilon, B_{e0}, \nu_0$ etc., and if the magnitude of $V$ turns out to be very small for given values of the parameters, it might be more convenient for experimental detection of circular polarization to calculate $C_l^{VT}$ rather than $C_l^{VV}$. The calculation of these parameters is beyond the main goal of this paper and will be considered elsewhere.

The fourth consideration is related with the possibility of generation of CMB polarization before decoupling epoch. Indeed,  milli-charged fermion vacuum polarization would generate CMB linear polarization even in the case when the CMB is initially unpolarized. In the case when $\Phi=\pi/2$ only the parameter $Q(T)$ is different from zero while in the case when $\Phi\neq \pi/2$ both parameters describing linear polarization $Q(T)$ and $U(T)$ are different from zero to second order in perturbation theory, while $V(T)$ is zero at this order. This effect is analogous to photon-pseudoscalar particle mixing in cosmic magnetic field which generates CMB polarization even before decoupling epoch time, see Ref. \cite{Ejlli:2016avx}.  In this situation, in addition to decay into milli-charged fermion, there would be also competing photon creation processes that would tend to push the system to thermal equilibrium state, so, the situation would be quite complicated, but it may be worth to pursue prior decoupling CMB polarization due to milli-charged fermion vacuum polarization.

 

 \vspace{2cm}

 {\bf{AKNOWLEDGMENTS}}:
 This work is partially supported by the Grant of the President of Russian Federation for the leading scientific Schools of Russian Federation, NSh-9022-2016.2 and by the top 5-100 program of Novosibirsk State University.

  \end{document}